\documentclass[traditabstract,epsf]{aa}

\PassOptionsToPackage{hyphens}{url}\usepackage[colorlinks = true, breaklinks]{hyperref}
\hypersetup{urlcolor=black,linkcolor=black,citecolor=black,colorlinks=true}
\usepackage{breakurl}
\usepackage{subfigure}
\usepackage{hyperref}
\usepackage{sistyle}
\usepackage{gensymb}

\DeclareTextSymbol{\degre}{OT1}{23}

\usepackage{amsmath}
\usepackage{graphicx}
\usepackage{epstopdf}
\usepackage{txfonts}
\usepackage{natbib}
\usepackage{verbatim}
\usepackage{epstopdf}
\usepackage[flushleft]{threeparttable}
\usepackage{threeparttable}

\begin{document}
   \title{Long-term activity and outburst of comet C/2013 A1 (Siding Spring) from narrow-band photometry and long-slit spectroscopy \thanks{Based on observations obtained at the ESO/VLT in the framework of program 93.C-0619.}}

\author{
 	  C.~Opitom\inst{1},
 	  A.~Guilbert-Lepoutre{$^2$},
      E.~Jehin\inst{1},
	  J.~Manfroid\inst{1},
	  D.~Hutsem\'ekers\inst{1},
	  M.~Gillon\inst{1},
	  P.~Magain\inst{1},
	  G.~Roberts-Borsani{$^3$},
	  O.~Witasse{$^4$}
       }

\offprints{cyrielle.opitom@ulg.ac.be}
\institute{ 
       	    $^1$ Institut d'Astrophysique et G\'eophysique, Universit\'{e} de Li\`{e}ge, all\'{e}e du 6 Ao\^{u}t 17, B-4000 Li\`{e}ge, Belgium\\  
       	    $^2$ CNRS - UTINAM UMR 6213, Avenue de l'Observatoire, 25000 Besançon, France \\
			$^3$Department of Physics and Astronomy, University College London, London NW1 2PS, United Kingdom       	\\
			$^4$ European Space Agency - ESTEC Kerplerlaan 1, 2201 AZ Noordwijk, Netherlands \\
       	  }
\date{Received date / accepted date}
\authorrunning{C. Opitom et al.}
\titlerunning{Comet C/2013 A1 (Siding Spring)}

  \abstract
  {In this paper, we present a unique data set of more than one year's worth of regular observations of comet C/2013 A1(Siding Spring) with TRAPPIST in Chile, along with low-resolution spectra obtained with the ESO/VLT FORS~2 instrument. The comet made a close approach to Mars on October 19, 2014 and was then observed by many space and ground-based telescopes. We followed the evolution of the OH, NH, CN, $\mathrm{C_3}$, and $\mathrm{C_2}$ production rates as well as the $Af\rho$ parameter as a proxy for the dust production. We detected an outburst two weeks after perihelion, with gas and dust production rates being multiplied by a factor five within a few days. By modelling the shape of the CN and $\mathrm{C_2}$ radial profiles, we determined that the outburst happened around  on November 10 around 15:30 UT ($\pm$ 5h) and measured a gas ejection velocity of $1.1\pm0.2$ km/s. We used a thermal evolution model to reproduce the activity pattern and outburst. Our results are consistent with the progressive formation of a dust mantle explaining the shallow dependence of gas production rates, which may be partially blown off during the outburst. We studied the evolution of gas composition, using various ratios such as CN/OH, $\mathrm{C_2}$/OH, or $\mathrm{C_3}$/OH, which showed little or no variation with heliocentric distance including at the time of the outburst. This indicates a relative level of homogeneity of the nucleus composition.
  }


   \maketitle



\section{Introduction}
\label{intro}

Nearly Isotropic Comets (NIC) are believed to have formed in the giant planet region, before being scattered onto their current orbits by planetary perturbations \citep{Dones2004}. The emplacement efficiency in the Oort Cloud is approximately 1-10\%, which means that most objects were lost to the interstellar medium in the early phases of the solar system evolution. Once a cometary nucleus has been placed in the Oort Cloud, it can return in the inner solar system due to gravitational perturbations from a passing star, molecular cloud or galactic tide. Since these comets have spent the last 3.5 to 4.5~Gyr at the edge of the solar system, barely bound to the Sun, in a collision-less environment at an equilibrium temperature of 10-20~K, they are possibly among the best preserved objects in the solar system. However, because they are rare and fast moving, they present very little opportunity to be thoroughly studied and remain very hard to investigate. In particular, we cannot expect to achieve the level of understanding we may get for Jupiter Family Comets (like comet 67P/Churyumov-Gerasimenko), which can be studied by in-situ space missions. 

Comet C/2013 A1 (Siding Spring) (hereafter Siding Spring) is a new NIC, discovered in January 2013 (McNaught et al. 2013) while at 7.2 au from the Sun. Early orbit determinations suggested a very close encounter with the planet Mars on October 19, 2014. This close encounter occurred at 134,000 km from the centre of Mars, at a relative speed of 56 km/s \citep{Kelley2014}. This close approach represented both a threat for orbiters around Mars and a tremendous opportunity to observe this comet from both ground- and Mars-based facilities. In particular, it allowed the study of the deposition of cometary dust into Mars’ atmosphere, with potential studying the dust composition. The SHARAD radar aboard Mars Reconnaissance Orbiter (MRO) recorded high values of its Total Electron Content on the nightside \citep{Restano2015}, while the MARSIS radar aboard Mars Express (MEX) detected a strong ionospheric layer just below 100km of altitude, consistent with a meteor shower produced by cometary dust particles \citep{Gurnett2015}. These observations were complemented by the MAVEN data, in which metallic ions, Na, Mg, Al, K, Ti, Cr, Mn, Fe, Co, Ni, Cu, and Zn, and possibly Si and Ca, were identified from ion mass and ultraviolet spectrometry \citep{Benna2015,Schneider2015}.

In this paper, we present observations of comet Siding Spring obtained over more than a year with the 60-cm TRAPPIST telescope, along with low-resolution spectra obtained by the ESO/VLT at 6 epochs between July 27 and September 28, 2014. These observations allow us to study the evolution of this comet activity from $\sim$5~au pre-perihelion, to $\sim$1.4~au post-perihelion, and to detect an outburst shortly after perihelion which we try to characterize by modelling the comet thermal evolution. Sections \ref{TRAPPIST} and \ref{VLT/FORS 2} present the observations and data reduction, while Sect. \ref{analysis} presents the simultaneous analysis of both TRAPPIST narrow-band images and VLT/FORS~2 low-resolution spectra. We present the evolution of the comet's dust  activity as it approached perihelion and shortly after in Sect. \ref{Dust}. We study the gas coma evolution and composition in Sect. \ref{rate}, and look for morphological features that could provide information about the position and number of active area on the nucleus in Sect. \ref{morph}. Finally, Sect. \ref{outburst} describes the characterisation of the outburst.

\section{Observations, data reduction and analysis}
\subsection{TRAPPIST}
\label{TRAPPIST}
\indent
TRAPPIST is a 60-cm robotic telescope installed in 2010 at La Silla observatory \citep{Jehin2011}. The telescope is equipped with a 2K$\times$2K thermoelectrically-cooled FLI Proline CCD camera with a field of view of 22\arcmin $\times$22\arcmin. We binned the pixels 2 by 2 and obtained a resulting plate scale of 1.302\arcsec/pixel. The telescope is equipped with a set of narrow-band filters designed for the observing campaign of comet Hale-Bopp \citep{Farnham2000}. These filters isolate the emission of OH (309.7 nm), NH (336.1 nm), CN (386.9 nm), $\mathrm{C_3}$ (406.3 nm), and $\mathrm{C_2}$ (513.5 nm), and emission free continuum regions at four wavelengths (UC at 344.9 nm, BC at 445.3 nm, GC at 525.9 nm, and RC at 713.3 nm). A set of B, V, Rc, and Ic Johnson-Cousin filters is also permanently mounted.

\indent
We started to observe the comet on September 20, 2013, when at 4.95~au from the Sun. At this time, it was too faint for gases to be detected with TRAPPIST, so we mainly observed with broad-band B, V, Rc, and Ic filters. We also got a few observations with the narrow-band continuum filters. Observations were performed once or twice a week, until the comet was lost due to the solar conjunction on April 7, 2014. We recovered the comet on May 21, 2014 and started to observe with both broad-band and narrow-band filters. We followed the comet until November 15, 2014, a few weeks after its perihelion passage. Exposure times range from 60 s to 240 s for the broad-band filters, and from 300 s to 900 s for the narrow-band filters.


\indent
Calibration followed standard procedures using frequently updated master bias, flat and dark frames. The removal of the sky contamination and the flux calibration were performed as described in \citet{Opitom2015}. Median radial profiles were extracted from each image and dust contamination was removed from gas radial profiles. OH, NH, CN, $\mathrm{C_3}$, and $\mathrm{C_2}$ flux were converted into column densities and we adjusted a Haser model \citep{Haser1957} on the profiles to derive the production rates. The model adjustment is performed around a physical distance of 10,000 km from the nucleus to avoid PSF and seeing effects around the optocenter and low signal-to-noise ratio at larger nucleocentric distances. The Haser model is not physically realistic as it assumes the single step photodissociation of parent species into daughter species in a spherically symmetric coma. However, it is widely used to compute gas production rates from optical comet observations and  allows to compare observations made by different observers, and also to compare comets between each others. We used scalelengths from \citet{AHearn1995} scaled as $r^2$, $r$ being the heliocentric distance. We chose tu use these wavelengths to allow an easy comparison with different data sets, especially the large data set from \citet{AHearn1995}. We used the observations with the narrow-band BC, GC, and RC filters and with the broad-band Rc filter to estimate the dust production. From the dust profiles, we derived the $Af\rho$ parameter, as first introduced by \citet{AHearn1984}. All $Af\rho$ values were corrected from the phase angle according to the phase function described by Schleicher\footnote{\url{http://asteroid.lowell.edu/comet/dustphase.html}}, which is a composite of two different phase functions from \citet{Schleicher1998} and \citet{Marcus2007}. The observational circumstances and production rates for each night are given in Table \ref{obstabtrappist}.

\begin{table*}[ht!]
\centering
\caption{$\mathrm{OH}$, $\mathrm{NH}$, $\mathrm{CN}$, $\mathrm{C_{3}}$, and $\mathrm{C_{2}}$ production rates and $Af\rho$ of comet C/2013 A1 (Siding Spring) from TRAPPIST observations.}
\begin{tabular}{p{2.05cm}p{0.35cm}p{0.40cm}p{0.40cm}p{1.55cm}p{1.55cm}p{1.55cm}p{1.55cm}p{1.55cm}p{1.55cm}}
  \hline
  \hline

  UT Date & $\mathrm{r}$ & $\Delta$ &  Sun PA & \multicolumn{5}{c}{Production rates ($10^{25}$ mol/s)} & $A(0) f\rho$   \\
   & (au) & (au) & ($\deg$) &Q(OH)  & Q(NH) & Q(CN) & Q($\mathrm{C_{3}}$) & Q($\mathrm{C_{2}})$ & ($10^{3}$ cm) \\
  \hline
  \hline
2013 Nov 16.52 & 4.42 & 3.82 & 336 &              &               &               &               &               & $1.54\pm0.13$ \\	
2014 Jan 25.09 & 3.73 & 3.70 & 70  &              &               &               &               &               & $1.46\pm0.10$ \\	
2014 Jan 29.09 & 3.69 & 3.71 & 74  &              &               &               &               &               & $1.48\pm0.08$ \\	
2014 Feb 04.06 & 3.63 & 3.73 & 79  &              &               &               &               &               & $1.46\pm0.08$ \\	
2014 Feb 12.05 & 3.55 & 3.76 & 86  &              &               &               &               &               & $1.50\pm0.13$ \\	
2014 Mar 26.01 & 3.13 & 3.76 & 127 &              &               &               &               &               & $1.06\pm0.08$ \\	
2014 Jun 02.42 & 2.43 & 2.93 & 230 &              &               &               &               &               & $1.03\pm0.05$ \\	
2014 Jun 03.42 & 2.42 & 2.91 & 209 &              &               & $1.10\pm0.11$ &               &               &               \\	
2014 Jun 17.40 & 2.27 & 2.60 & 220 &              &               & $1.20\pm0.11$ &               &               & $1.14\pm0.08$ \\
2014 Jun 20.44 & 2.24 & 2.53 & 222 &              &               &               &               &               & $1.19\pm0.06$ \\
2014 Jun 21.43 & 2.23 & 2.51 & 223 &              &               & $1.27\pm0.13$ &               &               & $1.16\pm0.04$ \\
2014 Jun 22.43 & 2.22 & 2.49 & 223 &              &               & $1.27\pm0.14$ &               &               &               \\
2014 Jun 24.42 & 2.20 & 2.44 & 225 &              &               & $1.26\pm0.09$ &               & $1.67\pm0.11$ & $1.15\pm0.03$ \\
2014 Jun 25.43 & 2.19 & 2.41 & 225 & 			  &               & $1.13\pm0.12$ &               & $1.51\pm0.14$ & $1.11\pm0.03$ \\
2014 Jun 26.43 & 2.18 & 2.39 & 226 &              &               &               &               &  			  &               \\
2014 Jun 30.42 & 2.15 & 2.29 & 228 &              &               & $1.42\pm0.15$ &               &  			  & $1.11\pm0.03$ \\
2014 Jul 01.42 & 2.14 & 2.26 & 229 & 			  &               & $1.32\pm0.08$ &               &               &   \\
2014 Jul 06.38 & 2.09 & 2.14 & 232 & 			  &               & $1.11\pm0.07$ & $0.58\pm0.03$ & $1.48\pm0.11$ & $1.12\pm0.06$ \\
2014 Jul 10.42 & 2.05 & 2.03 & 235 & $833\pm130$  &               & $1.52\pm0.09$ &               & $1.78\pm0.09$ & $1.18\pm0.03$ \\
2014 Jul 13.41 & 2.02 & 1.95 & 236 &              &               &               & $0.33\pm0.05$ &               & $1.12\pm0.06$ \\
2014 Jul 18.42 & 1.97 & 1.82 & 240 &              &               & $1.50\pm0.13$ & 			  & $1.38\pm0.11$ & $1.19\pm0.05$ \\
2014 Jul 21.42 & 1.94 & 1.74 & 242 & $735\pm100$  &               &               &               &               &               \\
2014 Jul 24.39 & 1.92 & 1.66 & 244 &              &               & $1.38\pm0.07$ & $0.38\pm0.03$ & $1.70\pm0.10$ & $1.15\pm0.02$ \\
2014 Jul 25.40 & 1.91 & 1.63 & 245 &              &               &               & $0.35\pm0.03$ & $1.68\pm0.07$ &  \\
2014 Aug 07.27 & 1.79 & 1.31 & 257 &              &               & $1.57\pm0.10$ &               & $2.00\pm0.21$ & $1.10\pm0.04$ \\
2014 Aug 10.34 & 1.77 & 1.24 & 261 & $863\pm106$  & $2.04\pm0.87$ & $1.51\pm0.09$ & $0.36\pm0.05$ & $1.77\pm0.15$ & $1.08\pm0.04$ \\
2014 Aug 17.39 & 1.71 & 1.09 & 274 & $830\pm132$  & $2.45\pm1.03$ & $1.57\pm0.11$ & $0.49\pm0.05$ & $1.91\pm0.12$ & $1.07\pm0.08$ \\
2014 Aug 20.35 & 1.69 & 1.04 & 281 & $808\pm143$  &               & $1.37\pm0.14$ & $0.35\pm0.05$ & $1.51\pm0.16$ & $1.11\pm0.05$ \\
2014 Sep 04.22 & 1.58 & 0.89 & 13  & $757\pm117$  & $3.49\pm0.65$ & $1.45\pm0.07$ &               & $2.04\pm0.13$ & $1.05\pm0.04$ \\
2014 Sep 06.98 & 1.56 & 0.89 & 37 & $712\pm129$   & $2.89\pm0.98$ & $1.52\pm0.11$ & $0.43\pm0.04$ & $1.87\pm0.13$ & $0.97\pm0.05$ \\
2014 Sep 07.31 & 1.56 & 0.89 & 39 & $780\pm154$   & 			  & $1.65\pm0.10$ &               & $2.04\pm0.13$ &               \\
2014 Sep 19.08 & 1.49 & 1.00 & 84 & $785\pm109$   &               & $1.36\pm0.06$ & $0.37\pm0.02$ & $1.56\pm0.11$ & $0.88\pm0.02$ \\
2014 Sep 24.08 & 1.47 & 1.08 & 90 & $824\pm111$   & $3.79\pm0.53$ & $1.47\pm0.08$ & $0.38\pm0.03$ & $1.83\pm0.19$ & $0.91\pm0.06$ \\
2014 Oct 06.05 & 1.43 & 1.33 & 95 & $879\pm112$   & $3.40\pm0.62$ & $1.56\pm0.07$ & $0.39\pm0.05$ & $1.63\pm0.17$ & $0.99\pm0.04$ \\
2014 Oct 13.05 & 1.41 & 1.48 & 95 & $1040\pm139$  & $4.19\pm0.48$ & $1.51\pm0.06$ & $0.40\pm0.03$ & $2.72\pm0.09$ & $0.99\pm0.03$ \\
2014 Oct 17.03 & 1.40 & 1.57 & 94 &               &  			  &               &  			  & $2.14\pm0.07$ &  \\
2014 Oct 19.04 & 1.40 & 1.61 & 93 &               & $3.74\pm0.52$ & $1.45\pm0.06$ &               &  		      & $1.02\pm0.03$ \\
2014 Oct 20.03 & 1.40 & 1.63 & 93 & $832\pm137$   & $4.08\pm0.67$ & $1.59\pm0.07$ & $0.41\pm0.04$ & $1.82\pm0.14$  & $1.06\pm0.03$ \\
2014 Oct 29.02 & 1.40 & 1.82 & 89 & $1040\pm210$  &               & $2.83\pm0.13$ &               & $2.67\pm0.15$ & $1.18\pm0.05$ \\
2014 Nov 07.02 & 1.41 & 2.00 & 84 &               &               & $5.60\pm0.33$ &               & $4.81\pm0.45$ &               $3.07\pm0.13$ \\
2014 Nov 11.02  & 1.42 & 2.07 & 80 &              &               & $22.1\pm1.4$  &               & $28.9\pm1.5$  & $9.21\pm0.42$ \\
2014 Nov 12.01 & 1.42 & 2.09 & 79 & $9010\pm3370$ &               & $14.7\pm1.2$  & $3.69\pm0.47$ & $25.5\pm1.7$  & $10.10\pm0.60$ \\
2014 Nov 13.02  & 1.43 & 2.10 & 78 &              &               & $10.1\pm0.9$  & $1.80\pm0.38$ & $14.0\pm1.4$  & $9.08\pm0.55$ \\
2014 Nov 14.02  & 1.43 & 2.12 & 77 &              &               & $7.84\pm0.92$ & $1.31\pm0.40$ & $12.2\pm1.4$  & $8.52\pm0.65$ \\
2014 Nov 15.01 & 1.43 & 2.13 & 76 &               &               & $6.77\pm0.79$ & $1.58\pm0.38$ & $9.90\pm1.39$ & $6.50\pm0.54$ \\
\hline
\hline
\end{tabular}
\label{obstabtrappist}
\tablefoot{
$r$ and $\Delta$ are respectively the heliocentric and geocentric distances (at 2 au one pixel represents about 2000 km). The date given in the first column is the mid-time of the observations.}
\end{table*}


\subsection{VLT/FORS~2}
\label{VLT/FORS 2}

\indent
Low-resolution spectra were acquired with the FORS~2 instrument \citep{Appenzeller1998} installed at UT1 of the Very Large Telescope of the European Southern Observatory. We obtained spectra of comet Siding Spring on six dates in 2014: Jul 7 ($r=1.91$ au), Aug 1 ($r=1.85$ au), Aug 29 ($r=1.62$ au), Aug 31 ($r=1.61$ au), Sep 19 ($r=1.49$ au) and Sep 28 ($r=1.45$ au). All observations were made using the grism 150I covering the 330-600~nm range. We chose a 6.8\arcmin -long and 1.3\arcsec -wide slit. The FORS~2 detector is composed of two 2K $\times$ 2K E2V CCDs (15 $\mu$m pixel size) separated by a 480 $\mu$m gap. The pixel scale is 0.25\arcsec/pixel in the spatial direction and 6.9$\angstrom$/pixel in the spectral direction. For each observation, 2 or 3 short exposures (from 30 to 180 s depending on the comet brightness and geocentric distance) ensured non-saturated observations of the nucleus and one long exposure (1500 s) allowed to get good signal-to-noise (SNR) ratio for the gas at larger nucleocentric distances. Except for the first observations, the slit was oriented perpendicular to the tail of the comet. Since the coma was filling the whole slit, we took exposures with a large offset every night to account for the sky background. We noticed a strong contamination from unknown origin in all September 28 spectra. This contamination was mostly located in the blue part of the spectra and prevented us from deriving reliable production rates. We thus discarded these observations from the dataset presented here.

\indent
We removed the bias, and then flat fielded and wavelength calibrated the spectra. The absolute flux calibration was made from spectroscopic standard stars observed the same night as the comet. The sky background was calibrated the same way as the scientific images. We measured the emission in the strongest sky lines from the background and comet images in order to properly scale the background and subtract it from the scientific images. This was done separately for each image and for each chip. The best way to account for the sky background would be to alternate comet and background expositions but there was not enough observing time available to apply this strategy. After the background subtraction, we binned the pixels by 10 in the spatial direction to increase the SNR. Each bin was then extracted separately. Observations of solar analogs (made with the same grism and slit width as the comet observations) were used to subtract the continuum from each 1D spectrum across the spatial direction. Spatial profiles of CN and $\mathrm{C_2}$ were obtained by summing every 1D spectra over the 383-390.5~nm and 486-521~nm wavelength ranges. We obtained two CN and $\mathrm{C_2}$ radial profiles (one on each side of the nucleus) for each exposure. We discarded the first point of each profile, containing the nucleus, to avoid PSF and seeing effects and also strong dust contribution difficult to remove close to the nucleus. The Haser model was adjusted on the profiles to derive CN and $\mathrm{C_2}$ production rates. The derived production rates and observational circumstances are given in Table \ref{obstabvlt}. The values in this Table are the mean values for each night, and the error bars are computed from the dispersion of the production rates from a single night.

\begin{table*}[ht!]
\centering
\caption{$\mathrm{CN}$ and $\mathrm{C_{2}}$ production rates of C/2013 A1 (Siding Spring) from VLT/FORS~2 observations}
\begin{tabular}{p{2.05cm}p{0.45cm}p{0.45cm}p{0.45cm}p{1.40cm}p{1.93cm}p{1.93cm}}
  \hline
  \hline  
  UT Date & $\mathrm{r}$ & $\Delta$ & Sun PA & Slit PA  &\multicolumn{2}{c}{Production rates ($10^{25}$ mol/s)} 
 \\
   & (au) & (au) & ($\deg$) & ($\deg$) & Q(CN) & Q($\mathrm{C_{2}})$ \\
   
  \hline
  \hline  
2014 Jul 25.38  & 1.91 & 1.65 & 245 & 206 & 1.33$\pm$0.34 & 1.29$\pm$0.18\\
2014 Aug 1.38   & 1.85 & 1.46 & 251 & 195 & 1.35$\pm$0.26 & 1.29$\pm$0.16\\
2014 Aug 29.17  & 1.62 & 0.92 & 320 & 256 & 1.42$\pm$0.26 & 1.41$\pm$0.04\\
2014 Aug 31.05  & 1.61 & 0.91 & 334 & 256 & 1.25$\pm$0.16 & 1.43$\pm$0.14\\
2014 Sep 19.01  & 1.49 & 1.00 & 84  & 180  & 1.22$\pm$0.10 & 1.65$\pm$0.10\\

\hline
\hline
\end{tabular}
\label{obstabvlt}
\tablefoot{
$r$ and $\Delta$ are respectively the heliocentric and geocentric distances (at 2 au one pixel represents about 400 km). The date given in the first column is the mid-time of the observations.}
\end{table*}


\subsection{Comparative analysis of TRAPPIST and FORS 2 data}
\label{analysis}

\begin{figure*}[ht!]
\centering
\subfigure{ \includegraphics[width=5.5cm]{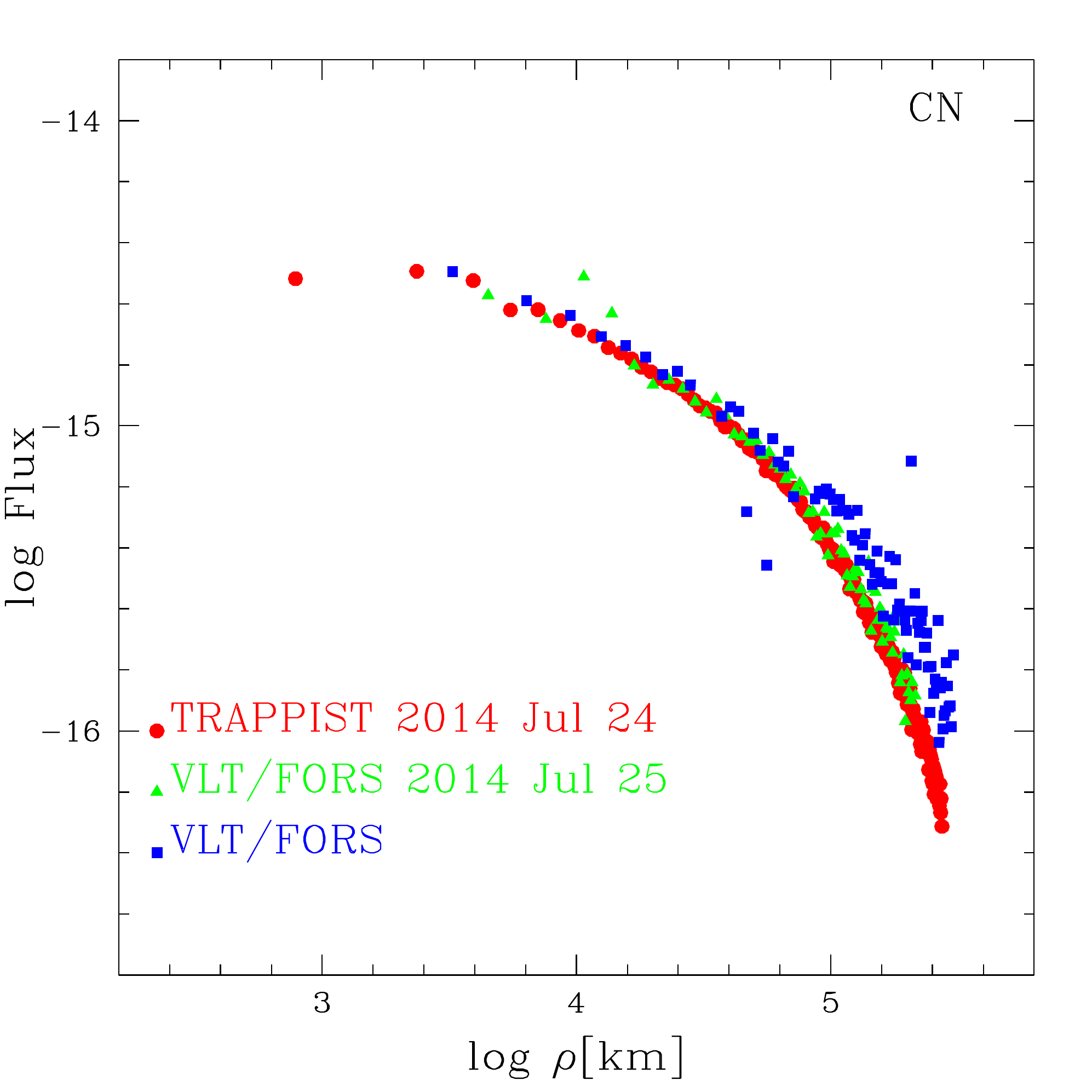}}
\hspace{1.5cm}
\subfigure{ \includegraphics[width=5.5cm]{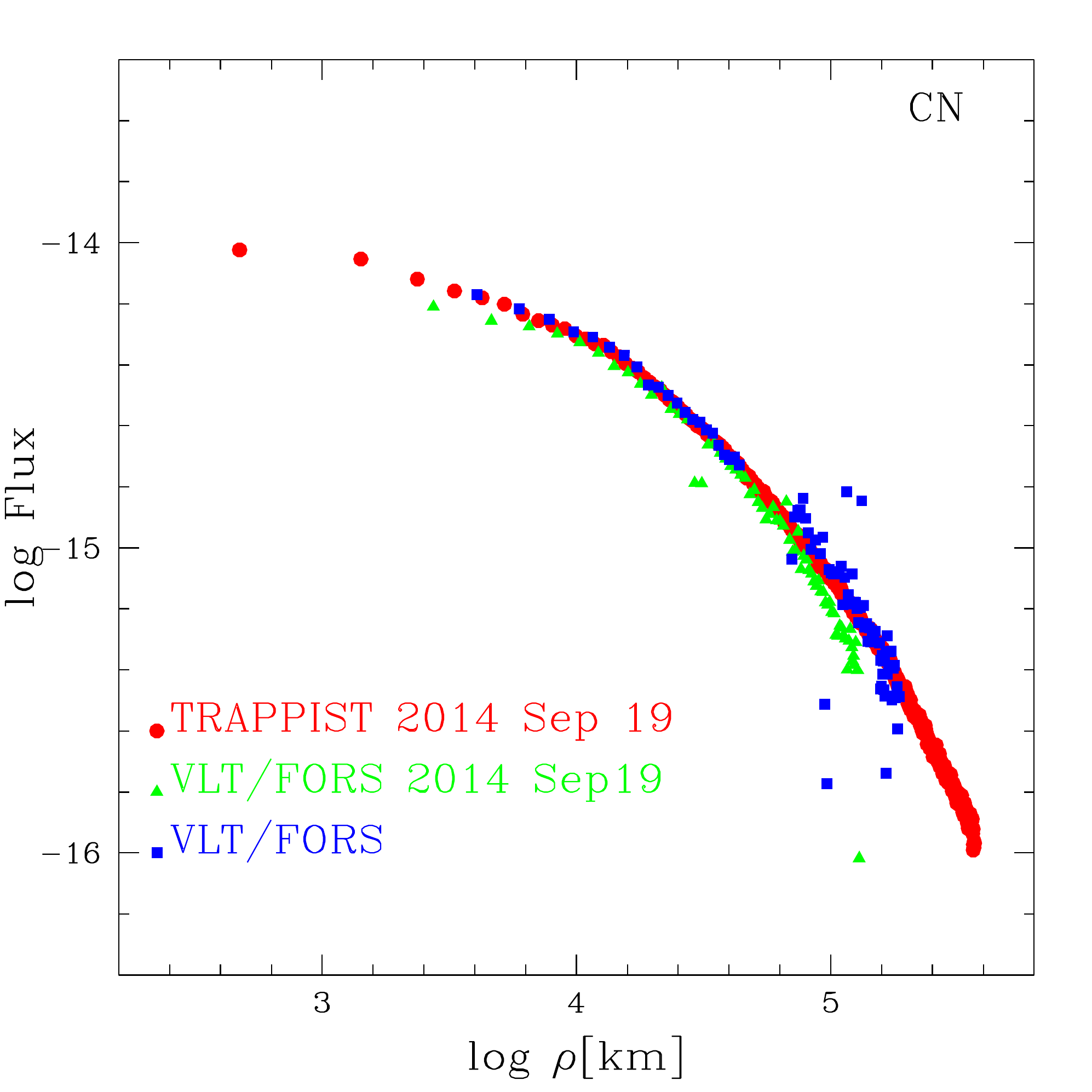}}
\subfigure{ \includegraphics[width=5.5cm]{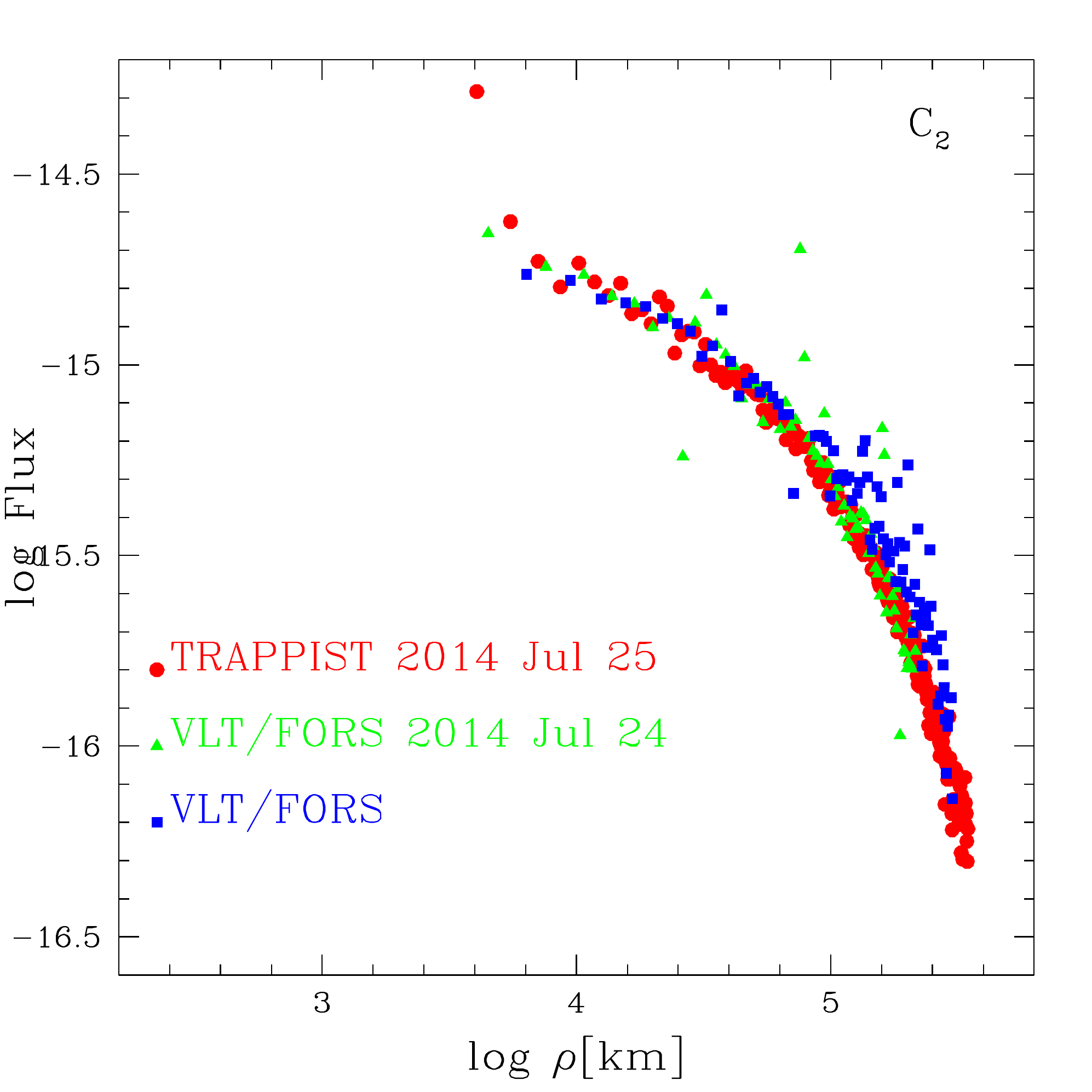}}
\hspace{1.5cm}
\subfigure{ \includegraphics[width=5.5cm]{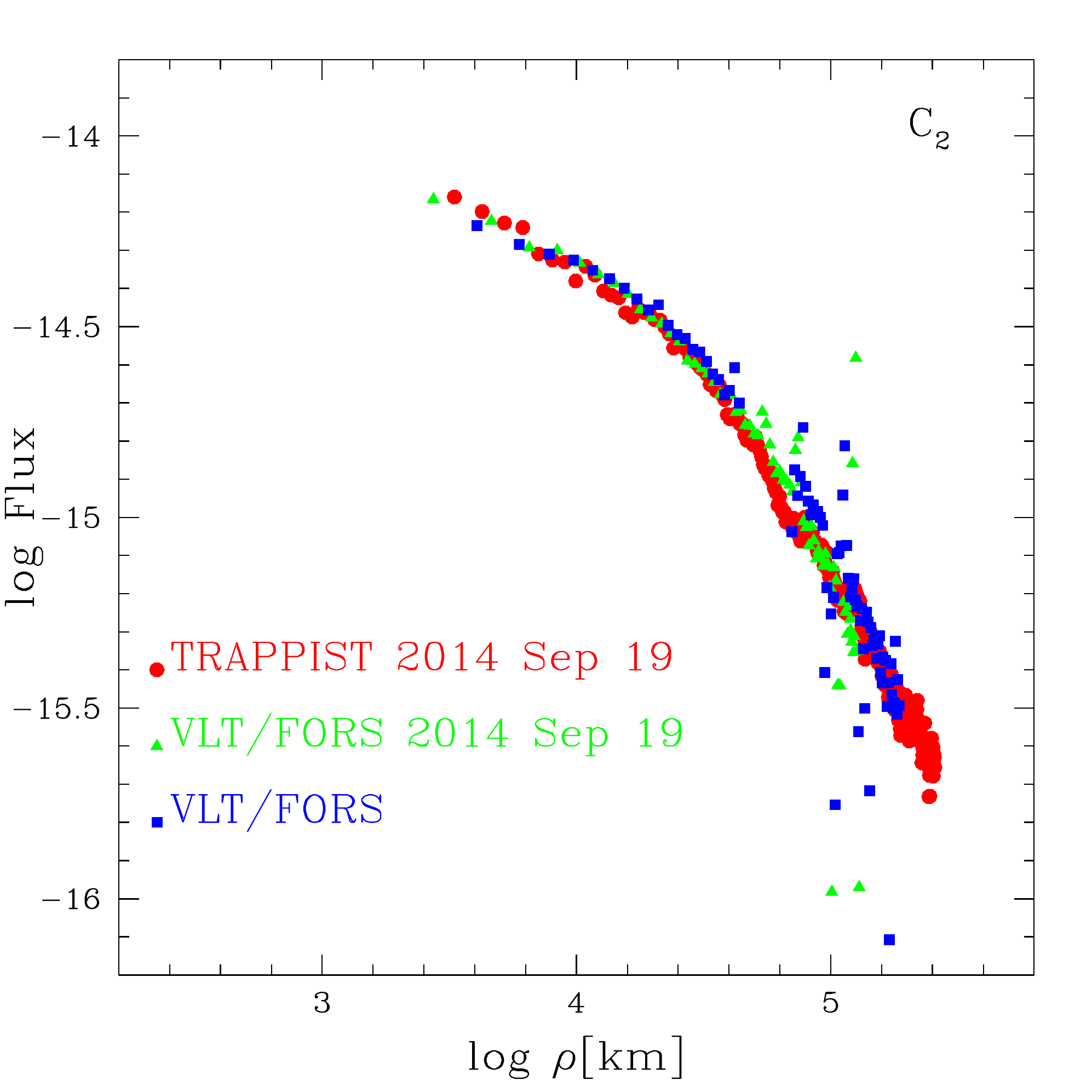}}
\caption{CN (top) and $\mathrm{C_2}$ (bottom) brightness profiles from Jul 25 and Sep 19, 2014. TRAPPIST data are represented with red circles and we overlaid the closest VLT data. Data extracted from both sides of the nucleus in the spectra are represented respectively with green triangles and blue squares.}
  \label{profilesCNC2}
\end{figure*}

\indent
We start our analysis by comparing the radial profiles and production rates obtained from TRAPPIST and FORS~2 data, to ensure they are consistent with each other. Indeed, it is usually difficult to compare data obtained with different instruments, and even more difficult to compare data obtained with different techniques. We only have almost simultaneous observations with FORS 2 and TRAPPIST for two dates, so we will use these observations to assess the agreement between both data sets. Figure \ref{profilesCNC2} shows CN and $\mathrm{C_2}$ radial brightness profiles for FORS~2 and the closest TRAPPIST observations. We first notice that the data dispersion is usually smaller in TRAPPIST profiles, especially at large nucleocentric distances, even though we binned the FORS~2 data by 10 pixels in the spatial direction to increase the SNR. The reason is that TRAPPIST profiles result from a median of the flux over the whole coma while FORS~2 profiles are only extracted over a 1.3\arcsec-wide slit. 

\indent
\indent
In the case of CN, TRAPPIST radial brightness profiles are in good agreement with the FORS~2 profiles, as shown in Fig. \ref{profilesCNC2}. The two FORS~2 profiles match each other, indicating that the coma is symmetric along the slit axis. The comparison of production rates derived from both telescopes (see Tables \ref{obstabtrappist} and \ref{obstabvlt}) shows that, within the error bars, production rates derived from FORS 2 observations fit within the trend of TRAPPIST production rates. Comparison of $\mathrm{C_2}$ fluxes between narrow-band images and low-resolution spectroscopy is more difficult than for the CN because of the extent of the band and the higher dust contamination. Despite this, our measurements of $\mathrm{C_2}$ radial profiles and production rates are also in good agreement between each other. Given that for two dates two months apart, TRAPPIST and FORS 2 radial profiles are a good match, and that production rates seem consistent with each others, we conclude that these observations can be merged within a single data set. We then analysed it all together while studying the evolution of the comet activity and coma composition. The results presented in the next Sections thus rely on both FORS~2 and TRAPPIST observations.

\section{Evolution of activity and coma}

\subsection{Dust}
\label{Dust}
\begin{figure}[ht!]
\centering
\includegraphics[width=7.5cm]{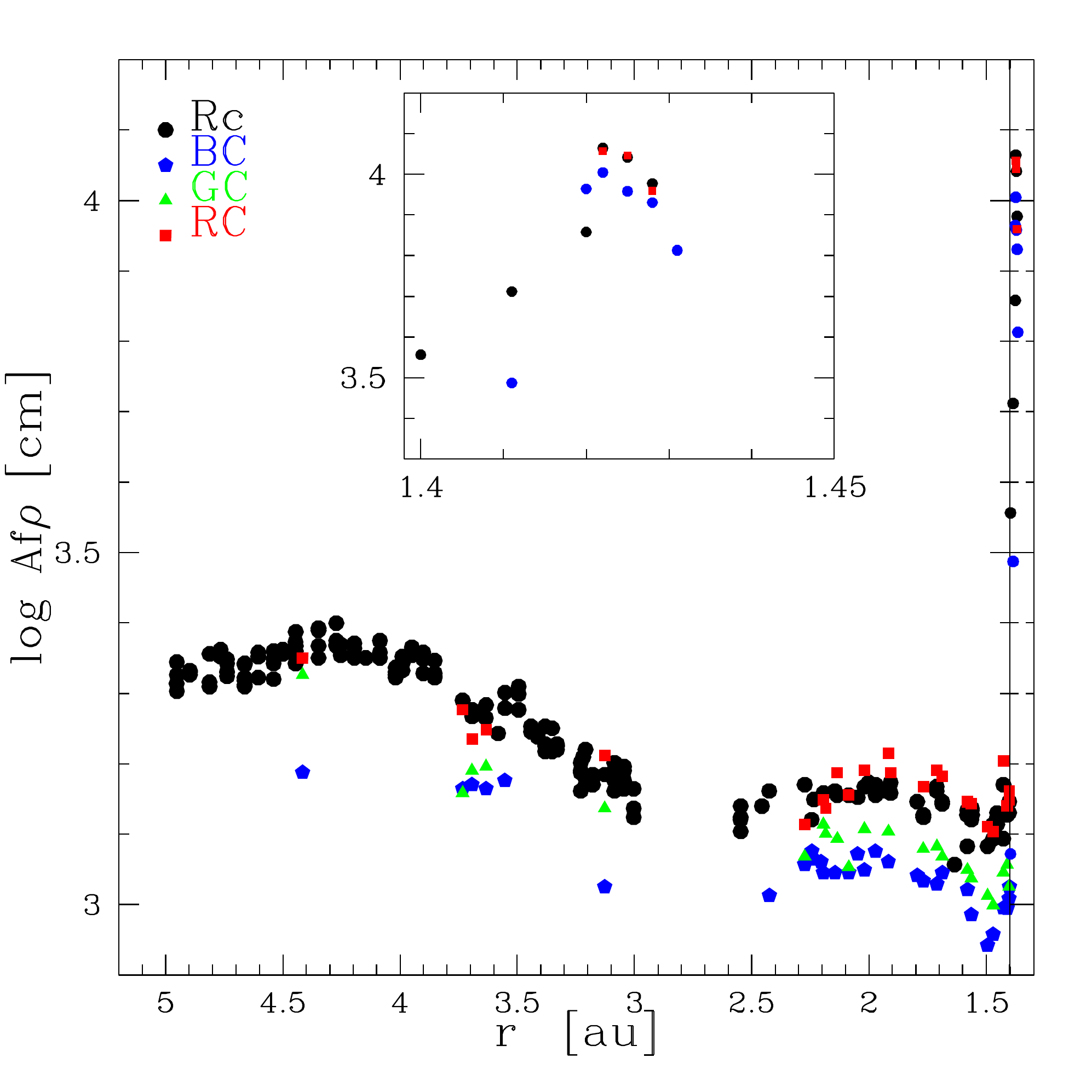}
\caption{The logarithm of the $A(0)f\rho$ corrected from phase angle effect as a function of the heliocentric distance $r$. The $A(0)f\rho$ was measured from Rc broad-band filter and from narrow-band BC, GC, and RC filters when the comet was bright. The vertical line indicates perihelion. We zoom on post-perihelion data.}
  \label{afrhoall}
\end{figure}

We observed comet Siding Spring during more than a year, almost uninterruptedly. Figure \ref{afrhoall} shows the evolution of the $A(0)f\rho$ corrected from phase angle effects, from almost 5~au pre-perihelion to more than two weeks after perihelion passage. The long-term trend observed is the same for all continuum filters. The major features are the following:
\begin{itemize}
\item The comet activity slowly rises as it approaches the Sun.
\item At approximately 4.3~au, the activity starts to decrease at a regular pace until the comet reaches 3~au. Because of the solar conjunction that prevented us from observing between Apr 7 and May 21, 2014, it is difficult to determine when the activity decrease stopped. 
\item From late May 2013 ($r = 2.5$~au) to mid July ($r=2.0$~au), the dust activity rises again, then goes through a standstill or a slow decrease period until the comet reached perihelion. 
\item After the perihelion passage on October 25 ($r=1.4$~au), the comet activity rises regularly. 
\item Between Nov 7 and Nov 11 ($r = 1.4$~au), the data show a sudden rise of the activity, consistent with a cometary outburst: both gas and dust production rates increase by a factor of 5 within a few days. Dust production peaks on Nov 12, then decreases regularly. 
\end{itemize}

\indent

\indent
\cite{Li2014} reported measurements of comet Siding Spring $Af\rho$ from observations performed with the Hubble Space Telescope. They measured a value of 2520~cm, 2120~cm, and 1720~cm at heliocentric distances of 5.58, 3.17, and 3.28~au respectively. These measurements are corrected from the phase angle effect using the same function as we used in this paper. From the closest TRAPPIST observations in the broad-band Rc filter, we measured $Af\rho$ values of $2230\pm30$, $1870\pm24$, and $1540\pm22$~cm at respectively 4.61, 3.73, and 3.33~au from the Sun. Our measurement are of the same order of magnitude but slightly lower than those reported by \cite{Li2014}. However, the $Af\rho$ values we report here are measured at a cometocentric distance of 10,000~km instead of 5,000~km for\cite{Li2014} values. Given that we observed a decrease of $Af\rho$ with cometocentric distance, our measurements are in fact consistent with \cite{Li2014}. Other $Af\rho$ values have been published by \cite{Stevenson2015} from NEOWISE  infrared observation, but any comparison between infrared and visible measurements is made difficult by color effects, and the different size of the field of view.

\begin{figure}[ht!]
\centering
\includegraphics[width=6.5cm]{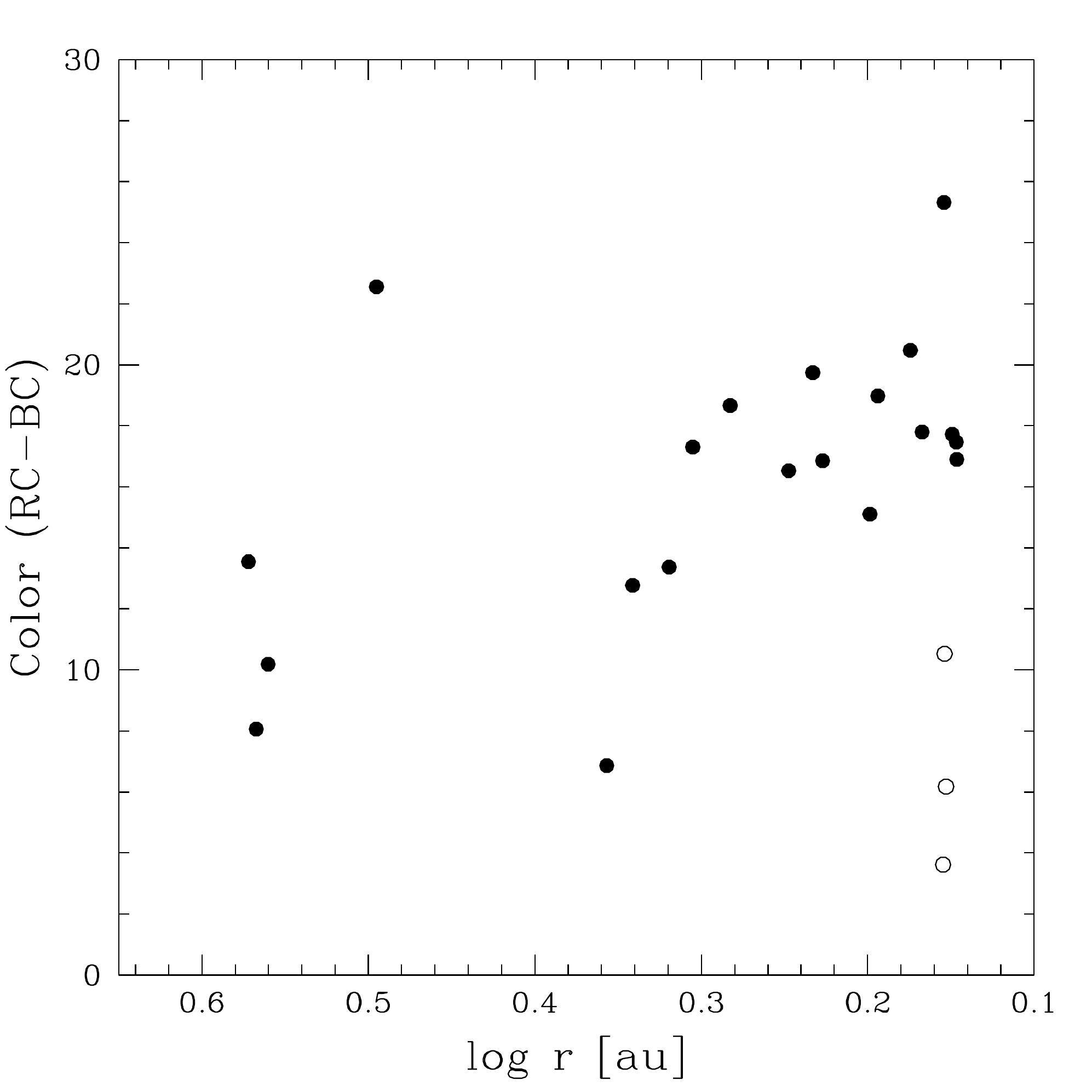}
\caption{Evolution of the dust color as a function of the heliocentric distance. Pre-perihelion values are represented with full circles and post-perihelion values with open circles.}
  \label{dustcolor}
\end{figure}

\indent
Taking advantage of our regular observations of the dust continuum through RC and BC narrow-band filters during the same night, we were able to study the dust color and its evolution with heliocentric distance. The dust color is usually computed as the normalized gradient of $A(\theta) f\rho$ from two continuum filters: 
\begin{equation}
\text{color} [\lambda _{1},\lambda _{2}]=\frac{Af\rho _{1}-Af\rho _{2}}{\lambda _{1}-\lambda _{2}}\frac{2000}{Af\rho _{1}+Af\rho _{2}}
\end{equation}
The normalized reflectivity gradient is expressed as the percentage of reddening by 1,000~$\angstrom$. Fig. \ref{dustcolor} shows the evolution of the dust color as a function of heliocentric distance. Because of the large dispersion, it is impossible to draw a clear trend with the heliocentric distance. Our measurements are slightly higher than those published by \citep{Li2014} (6\%/1,000 $\angstrom$ at 3.8~au), but consistent if we consider the large dispersion in the data. \citeauthor{Li2014} reported a reddening of the dust between 4.6 and 3.3~au, which they imputed to icy grains sublimating in the coma. We only have a few color measurements after perihelion, so it is difficult to assess the effect of the outburst on the dust color, even though we may be seeing a bluer dust at that time.


\subsection{Gas composition}
\label{rate}

\begin{figure*}[ht!]
\centering
\subfigure{ \includegraphics[width=6.5cm]{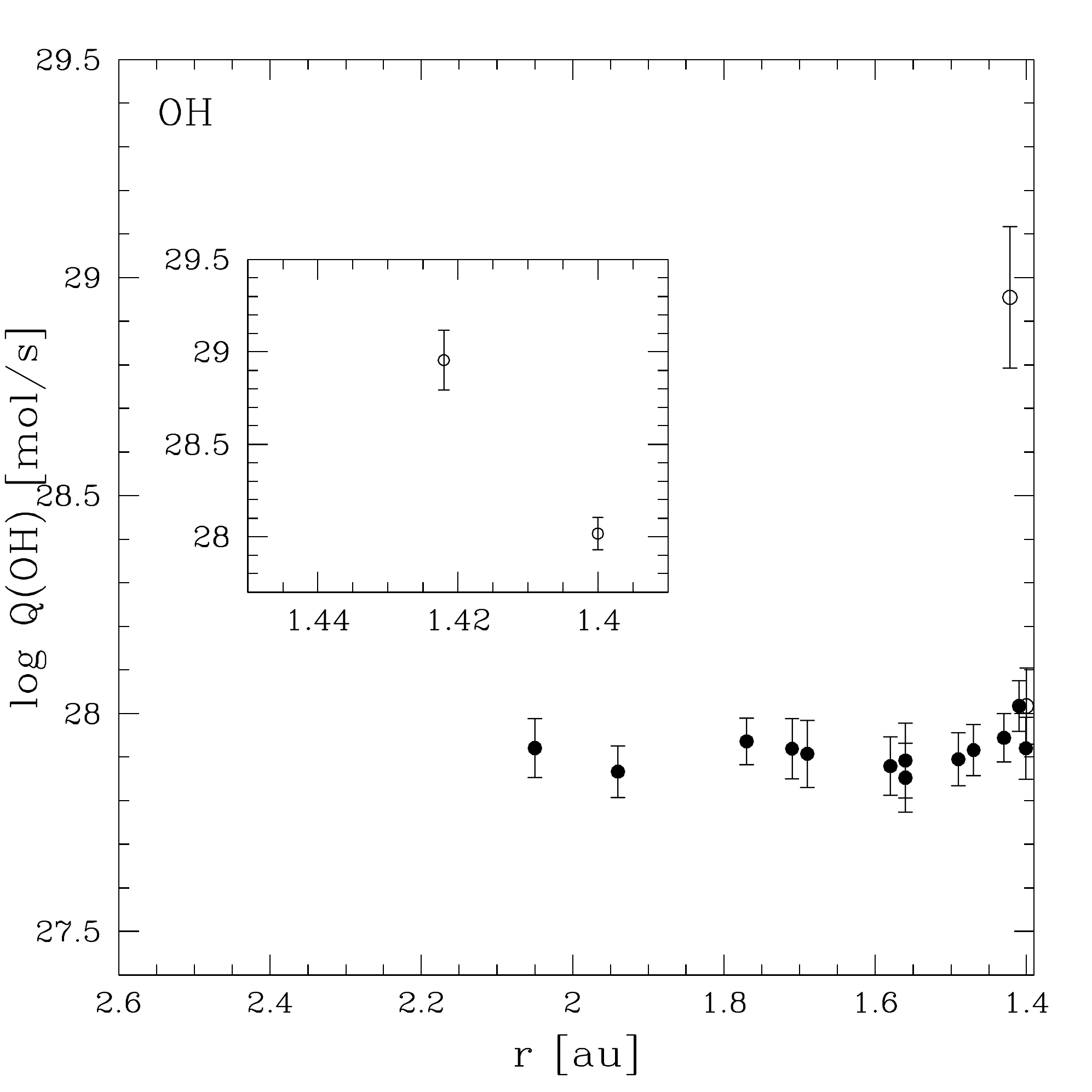}}
\hspace{1.5cm}
\subfigure{ \includegraphics[width=6.5cm]{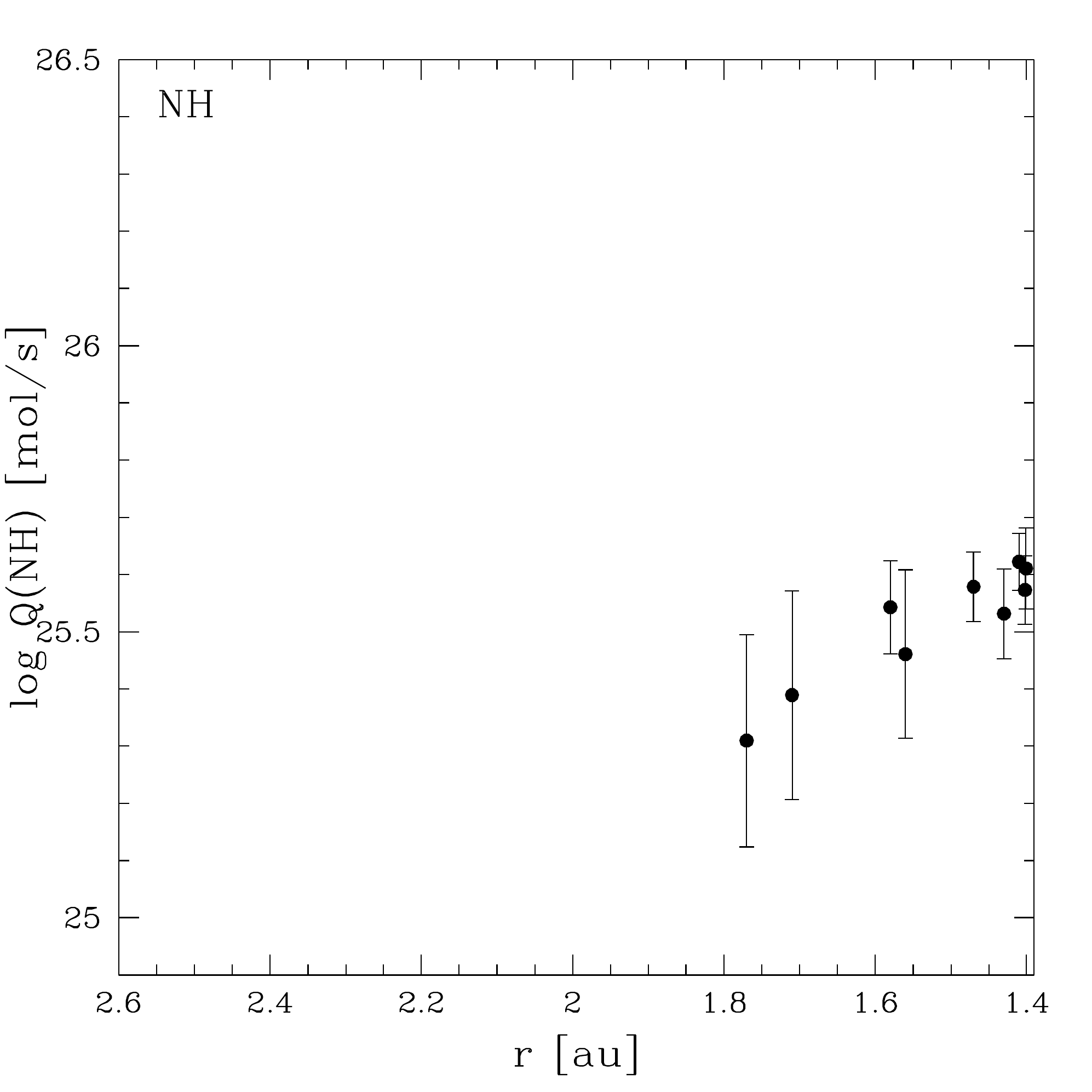}}
\subfigure{ \includegraphics[width=6.5cm]{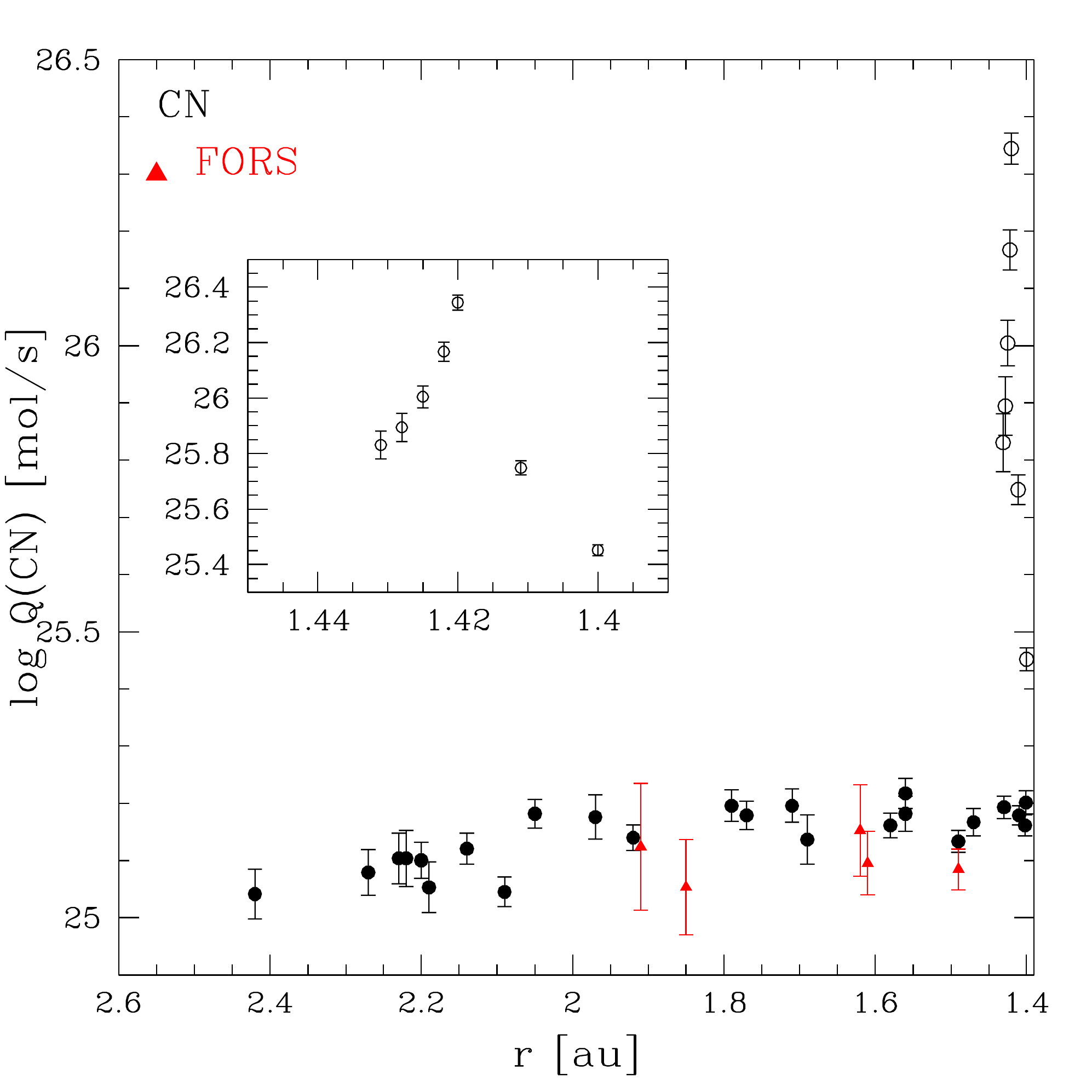}}
\hspace{1.5cm}
\subfigure{ \includegraphics[width=6.5cm]{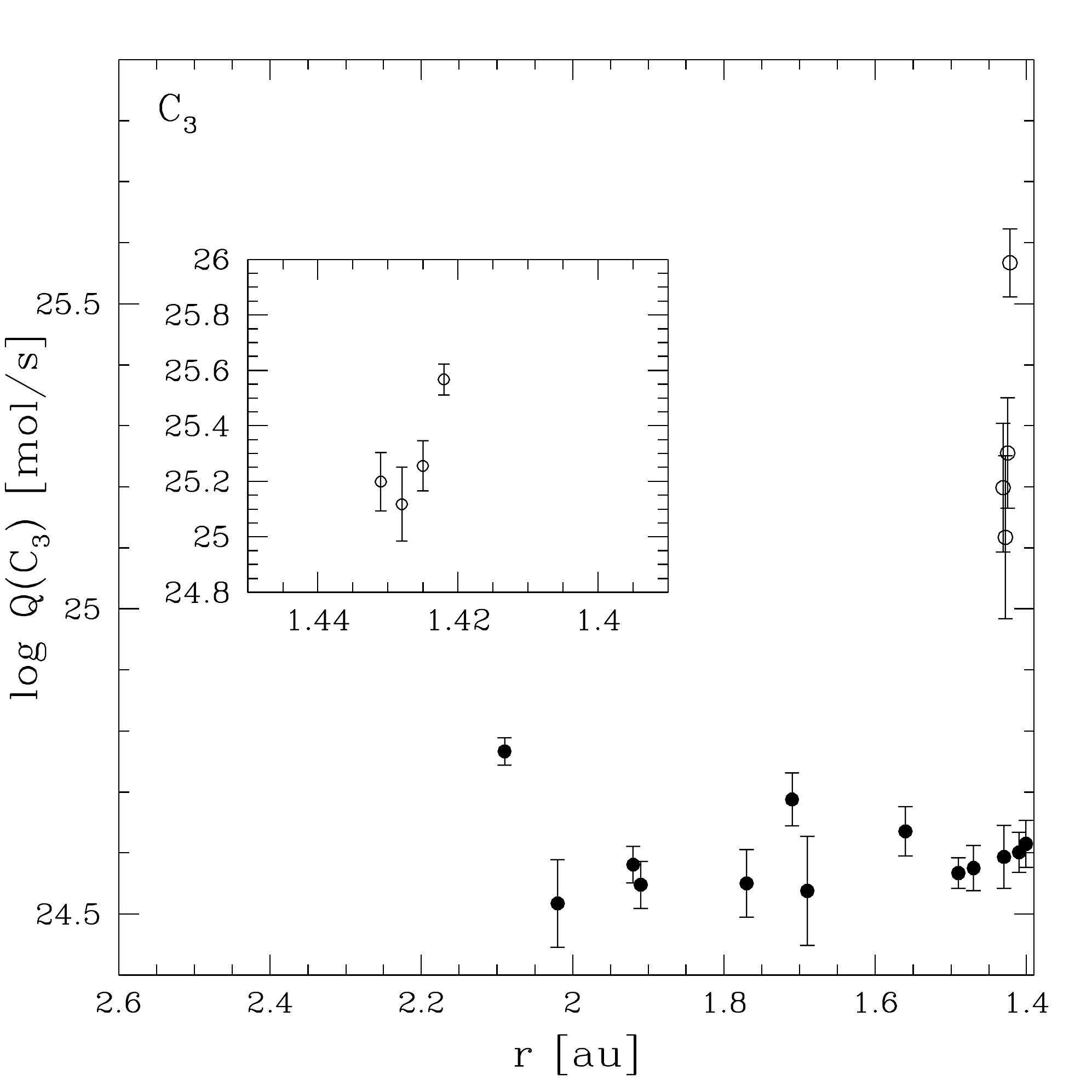}}
\subfigure{ \includegraphics[width=6.5cm]{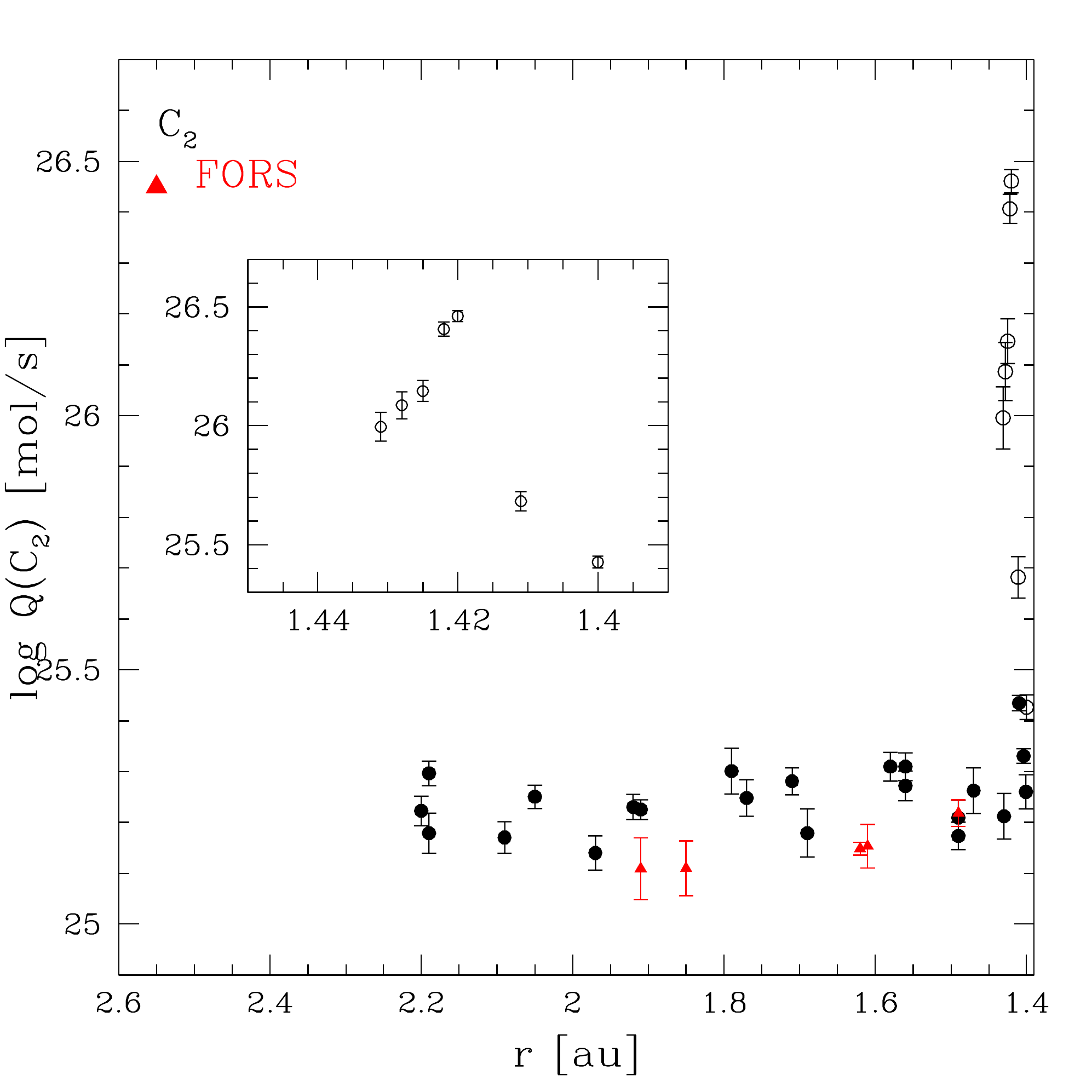}}
\hspace{1.5cm}
\subfigure{ \includegraphics[width=6.5cm]{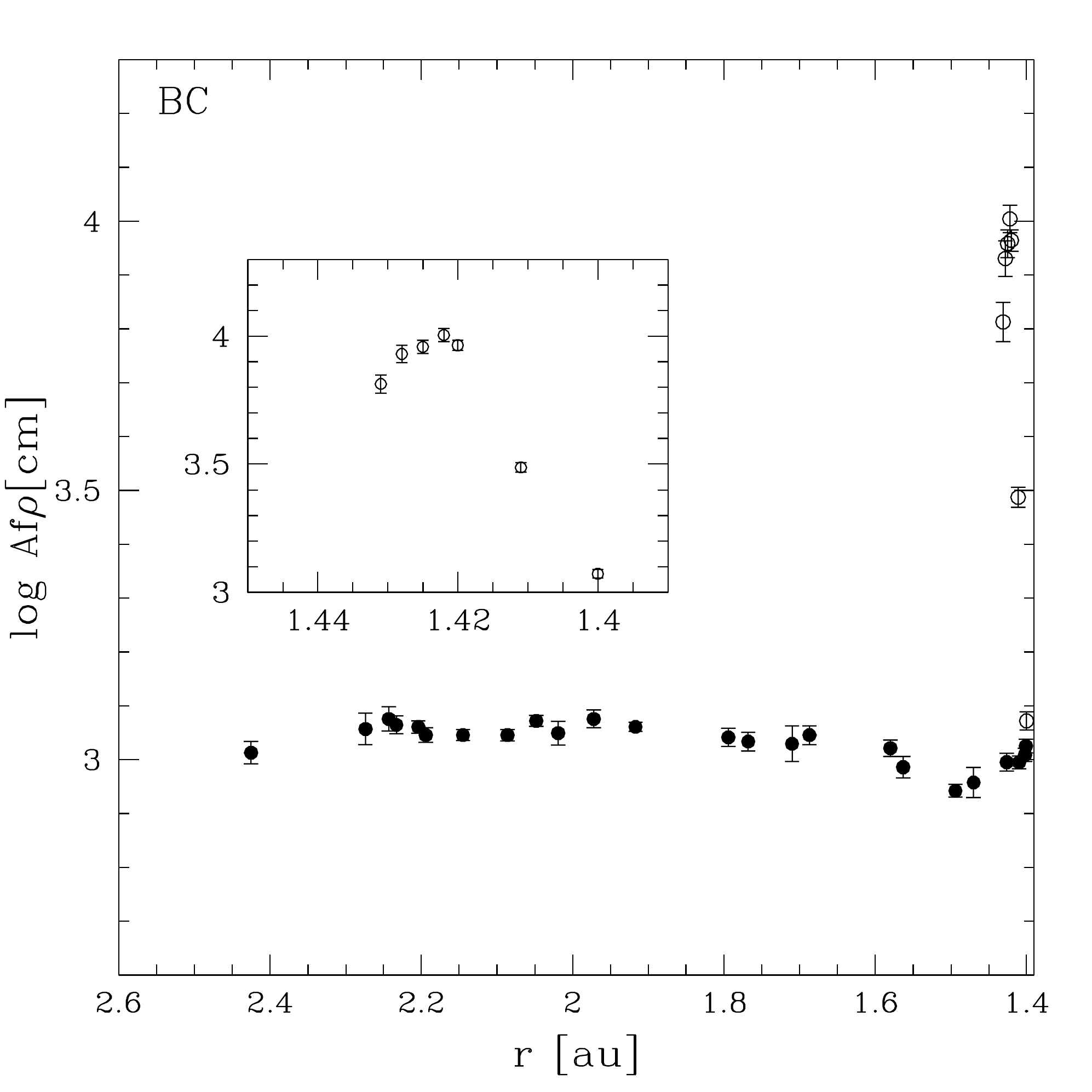}}
\caption{OH, NH, CN, $\mathrm{C_3}$, and $\mathrm{C_2}$ production rates and $A(0)f\rho$ as a function of the heliocentric distance ($r$). Pre-perihelion data are represented with filled symbols and post-perihelion data with open symbols. We zoom on post-perihelion data. For CN and C2, spectroscopic determination from FORS~2 have been added.}
  \label{productionrates}
\end{figure*}

\indent
Figure \ref{productionrates} shows the evolution of the gas production rates and $A(0)f\rho$ with the heliocentric distance from 2.43~au pre-perihelion  until after perihelion.  Fig. \ref{productionrates} and Table \ref{obstabtrappist} reveal there is no clear increase of most gas production rates while the comet approaches the Sun. They only slightly vary between 2.43~au and perihelion, which may seem surprising. The NH production rates are the only ones that increase significantly while the comet approaches the Sun. However, the uncertainties of the NH production rates are larger given the low SNR. The rise of the activity after perihelion followed by the outburst is also clearly visible in Fig. \ref{productionrates}. Gas production rates are multiplied by a factor 5 approximately between November 7 and November 11, 2014. They peak on November 11 and decrease during the following days. This outburst is further analysed in Section \ref{outburst}.

\indent
Water production rates were reported by \cite{Bodewits2015} from observations with the UltraViolet-Optical Telescope on board Swift between November 2013 and October 2014. They measured $\mathrm{H_{2}O}$ production rates of $11.1\pm1.0$ $10^{27}$, $12.1\pm0.27$ $10^{27}$, $13.0\pm0.45$ $10^{27}$, $12.5\pm0.62$ $10^{27}$, and $17.2\pm0.5$ $10^{27}$~mol/s on July 09, August 19, September 18, October 13, and October 23, 2014 respectively. They could not detect water until the comet reached 2.46~au pre-perihelion. Then they observed an increase of water production rates and active area between 2.46 and 2.0~au, followed by a plateau of water production rate along with a decrease of active area between 2~au and perihelion.
The closest TRAPPIST observations allowed us to derive equivalent water production rates ($Q(\mathrm{H_{2}O})=1.361r^{-0.5}Q(\mathrm{OH})$, \citealt{Cochran1993}) of $8.3\pm1.3$ $10^{27}$, $8.5\pm1.5$ $10^{27}$, $8.75\pm1.2$ $10^{27}$, $11.9\pm1.6$ $10^{27}$, and $9.6\pm1.6$ $10^{27}$ for observations made respectively on July 10, August 20, September 19, October 13, and October 20, 2014. Most of these observations are consistent within the error bars with values reported by \cite{Bodewits2015}. The only large discrepancy is between the $\mathrm{H_{2}O}$ production rates measured by \cite{Bodewits2015} on October 23, and our observations performed on October 20. The origin of this discrepancy is not clear, even though the production rate we report for October 20 is lower compared to contemporary observation in Table \ref{obstabtrappist}.

\begin{figure}[htb!]
\centering
\includegraphics[trim=0cm 0cm 8.8cm 0cm,width=8.0cm]{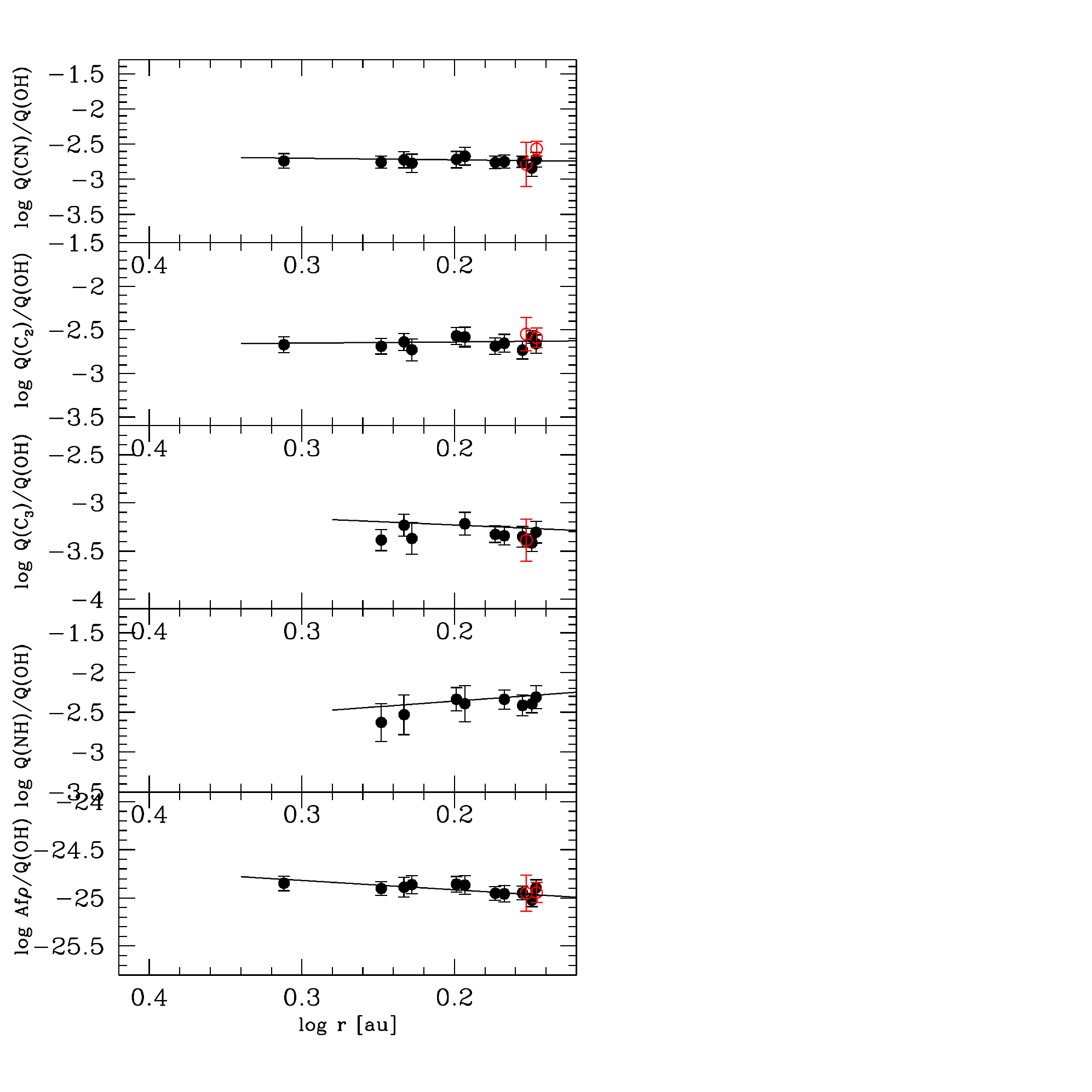}
\caption{Ratio of NH, CN, $\mathrm{C_3}$, and $\mathrm{C_2}$ production rates and the $A(0)f\rho$ to OH production rate as a function of the heliocentric distance. Pre-perihelion values are represented with filled symbols and post-perihelion values with open red symbols. Full lines represent linear fits of the ratios variation with the heliocentric distance pre-perihelion.}
  \label{compOH}
\end{figure}

\indent
We computed several production rate ratios in order to study the coma composition. Comet Siding Spring is a typical comet in terms of $\mathrm{C_2}$/CN and $\mathrm{C_3}$/CN ratios, as defined by \cite{AHearn1995}. The evolution of NH, CN, $\mathrm{C_3}$, $\mathrm{C_2}$ production rates, and the  $A(0)f\rho$ relative to OH, are represented on Fig. \ref{compOH}. A linear fit is adopted for each ratio, in order to assess the evolution of the coma composition more easily. The $\mathrm{C_2}$/OH, $\mathrm{C_3}$/OH and CN/OH ratios do not significantly vary with the heliocentric distance. To the contrary, the NH/OH ratio is increasing with decreasing heliocentric distance though. However, NH scalelengths and their dependence with heliocentric distance are poorly known, which makes it difficult to interpret the trend in the evolution of the NH/OH ratio. The bottom part of Fig. \ref{compOH} shows the evolution of the dust-to-gas ratio. The slope of the linear fit is significantly larger than zero, meaning that the coma is becoming less dusty as it approaches the Sun. Although we only have a few measurements of these ratios during the outburst, no significant variation of the coma composition at that particular time can be pointed out.

\subsection{Morphology}
\label{morph}

\begin{figure*}[ht!]
\centering
\subfigure{ \includegraphics[width=5.5cm]{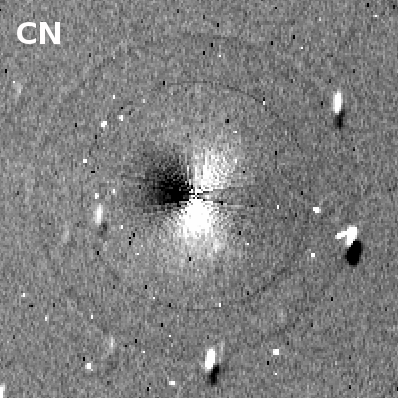}}
\hspace{1.5cm}
\subfigure{ \includegraphics[width=5.5cm]{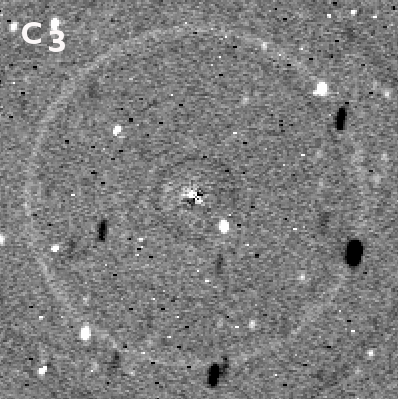}}
\subfigure{ \includegraphics[width=5.5cm]{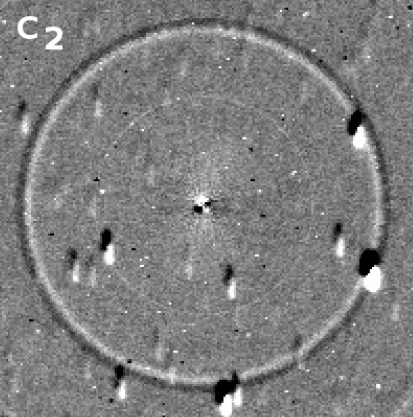}}
\hspace{1.5cm}
\subfigure{ \includegraphics[width=5.5cm]{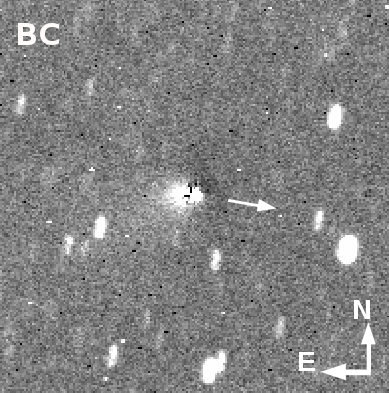}}
\caption{C/2013 A1 (Siding Spring) CN, $\mathrm{C_3}$, $\mathrm{C_2}$, and dust from BC filter images from November 11 and 12 processed by subtracting an azimuthal median profile. All images are oriented with North up and East left. The field of view is 4.3\arcmin $\times$ 4.3\arcmin . The arrow indicates the direction of the Sun. The circles in the images are artefacts due to a bright star in the field of view.}
  \label{jets}
\end{figure*}

\indent
In this section, we discuss the morphology of the coma and its evolution during our observations. We subtracted an azimuthal median profile from every image, in order to enhance the contrast between the average coma and the morphological features. Except for  CN, we only detect weak or no features. We show an example of CN, $\mathrm{C_3}$, $\mathrm{C_2}$, and dust from BC filter images in Fig.\ref{jets}. These images have been taken one day after the outburst on Nov 11, 2014 (but on Nov 12 for $\mathrm{C_3}$). The features visible in Fig. \ref{jets} were already there in August and September images but they were not as bright and contrasted. We do not observe any particular change of the coma morphology during the observations, or at the time of the outburst, except for a change of the features position caused by a variation of the viewing geometry. 

\indent
The BC enhanced image shows two small jets in opposite directions, different from the tail direction. Some of the BC images (not shown here) also display an enhancement of the coma in the tail direction, which is barely visible in Fig. \ref{jets}. The position of the dust jets is difficult to determine precisely, given their small spatial extension. The $\mathrm{C_3}$ and $\mathrm{C_2}$ enhanced images are very similar to each other. They both display two short jets in opposite directions. Even though it is difficult to determine their precise position, they appear close to the dust jets we identified before. However, they are weaker than the dust jets, and are not detected in every images before the outburst. Finally, the CN morphology is completely different: two broader and more extended fans can be observed around PA 185$\degree$ and PA 330$\degree$.

\section{Outburst}
\label{outburst}
\subsection{Observational characterization}

\begin{figure*}[ht!]
\centering
\subfigure{ \includegraphics[width=5.5cm]{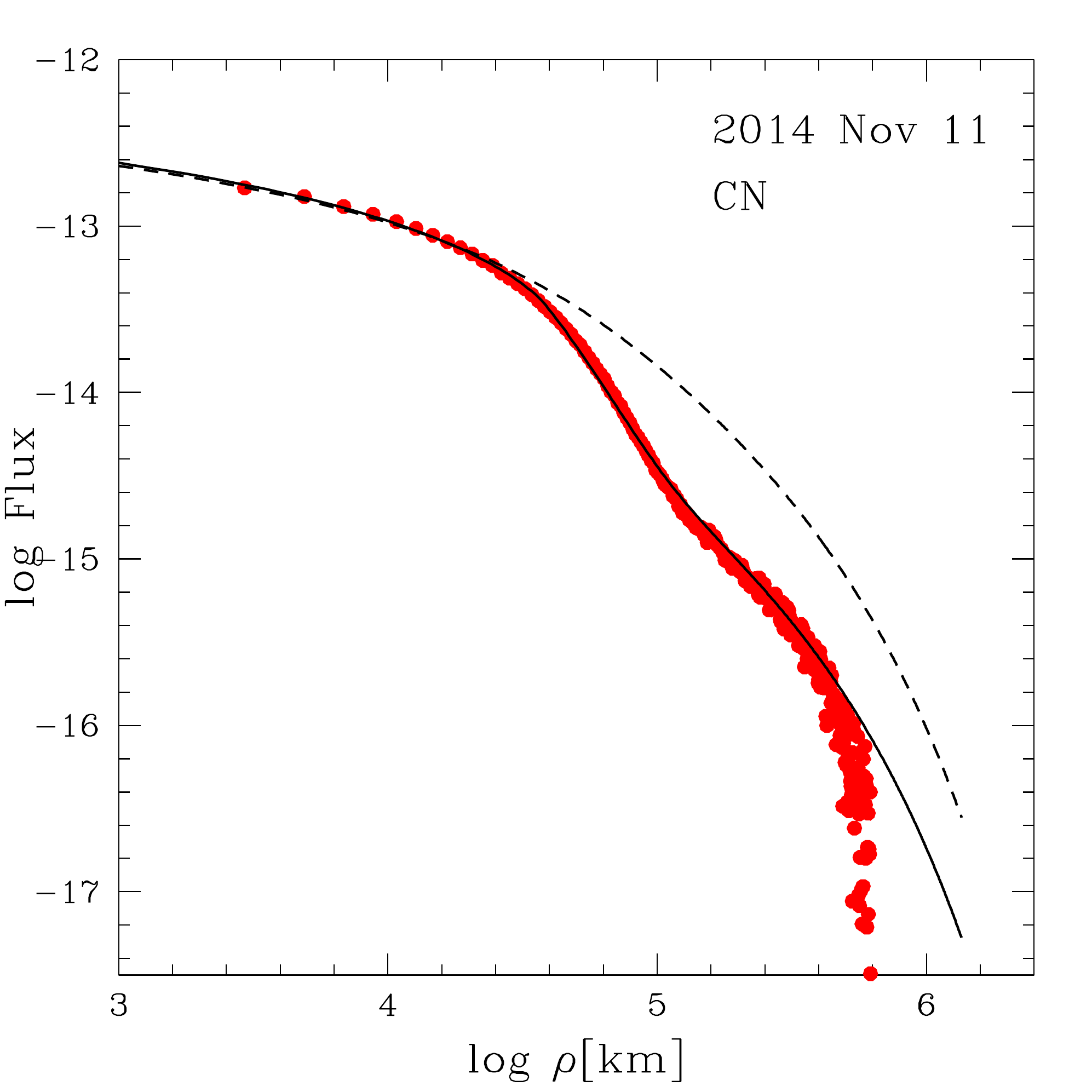}}
\hspace{0.5cm}
\subfigure{ \includegraphics[width=5.5cm]{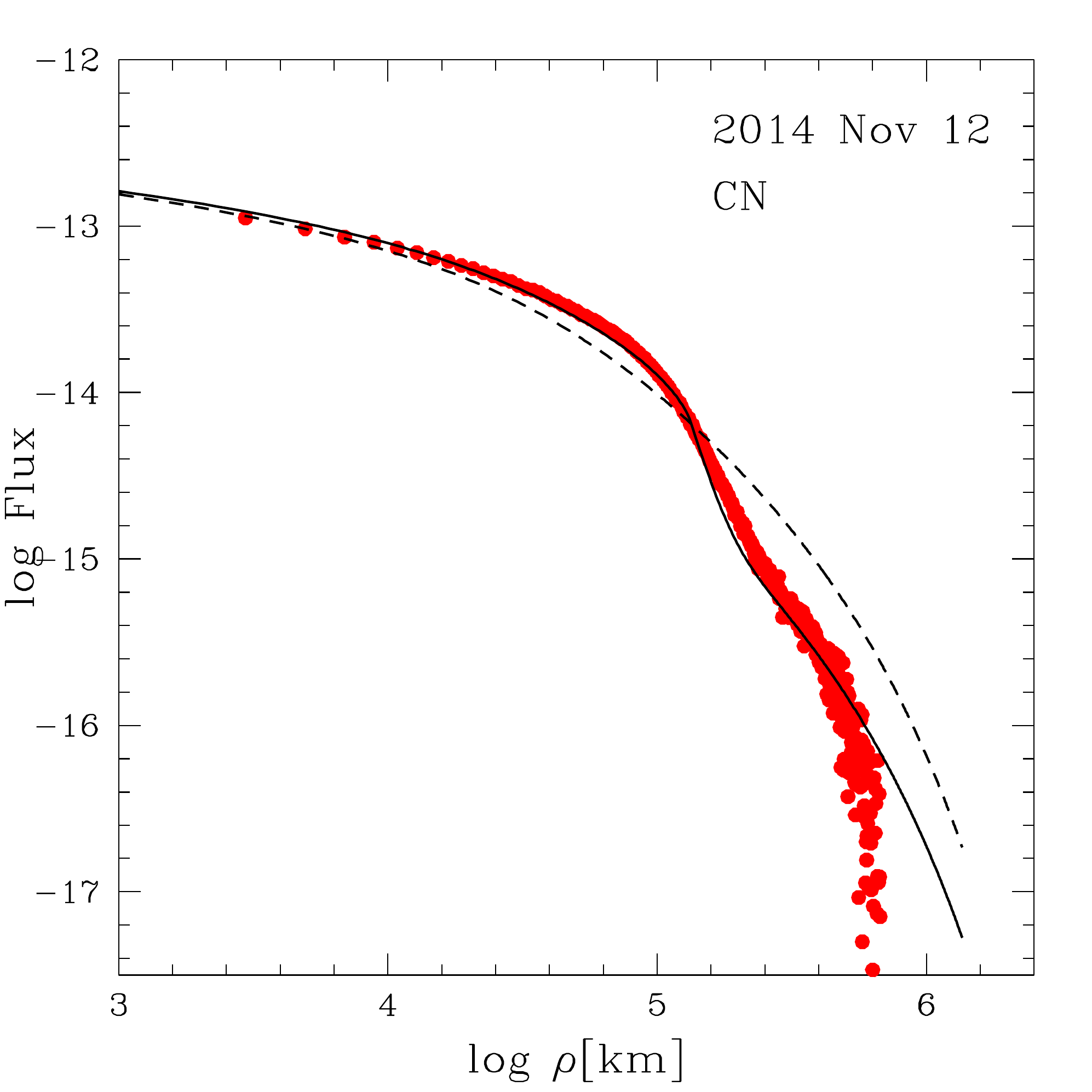}}
\hspace{0.5cm}
\subfigure{ \includegraphics[width=5.5cm]{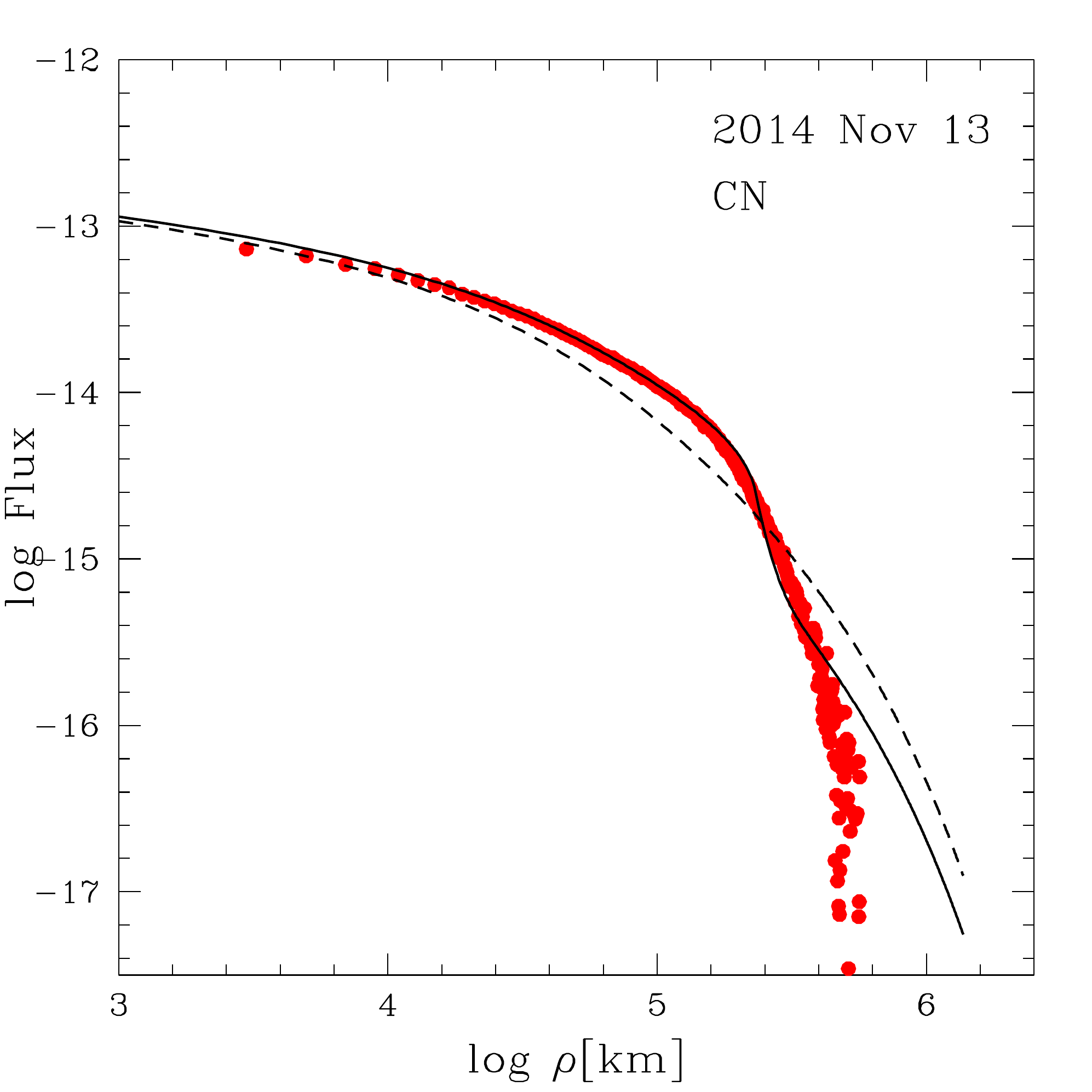}}
\subfigure{ \includegraphics[width=5.5cm]{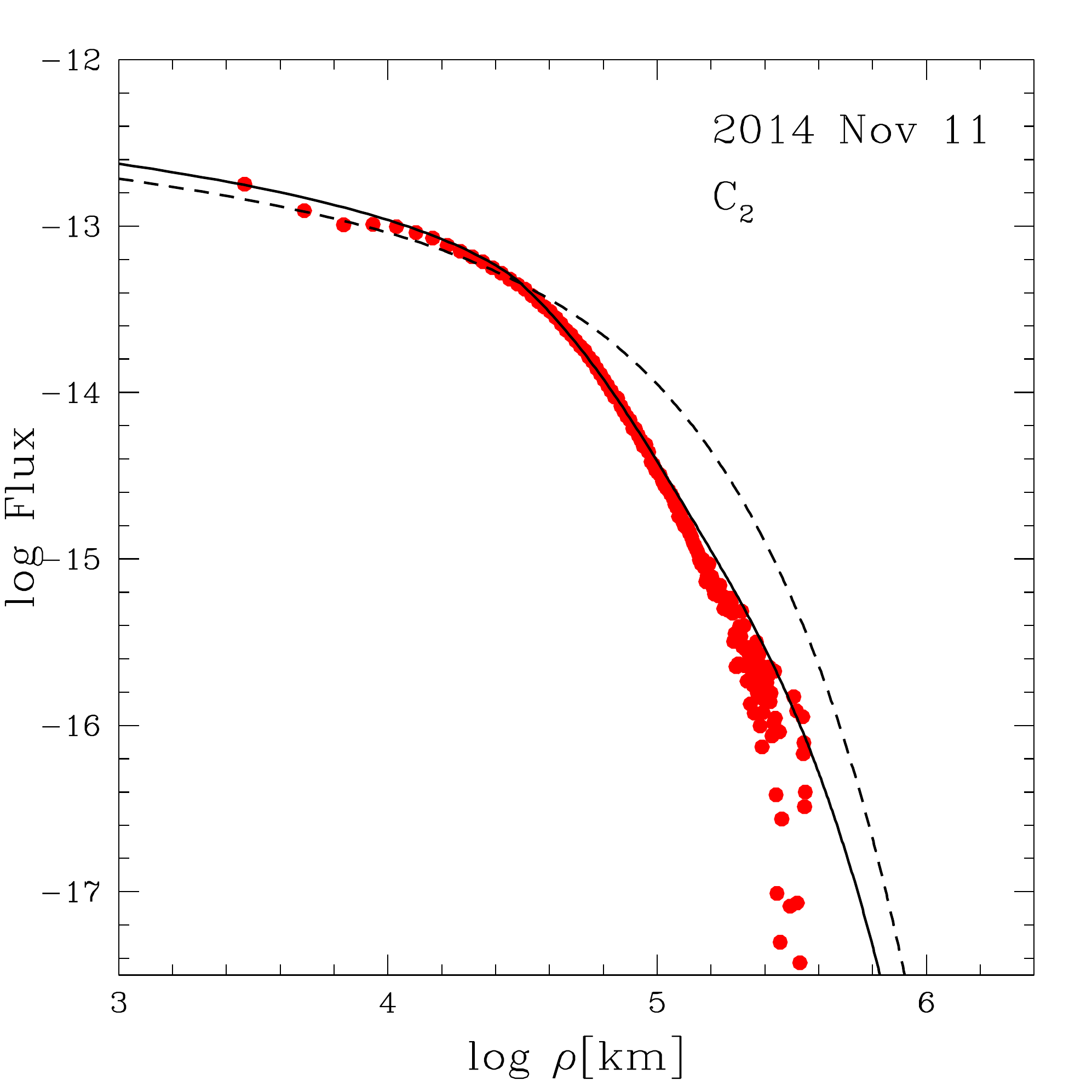}}
\hspace{0.5cm}
\subfigure{ \includegraphics[width=5.5cm]{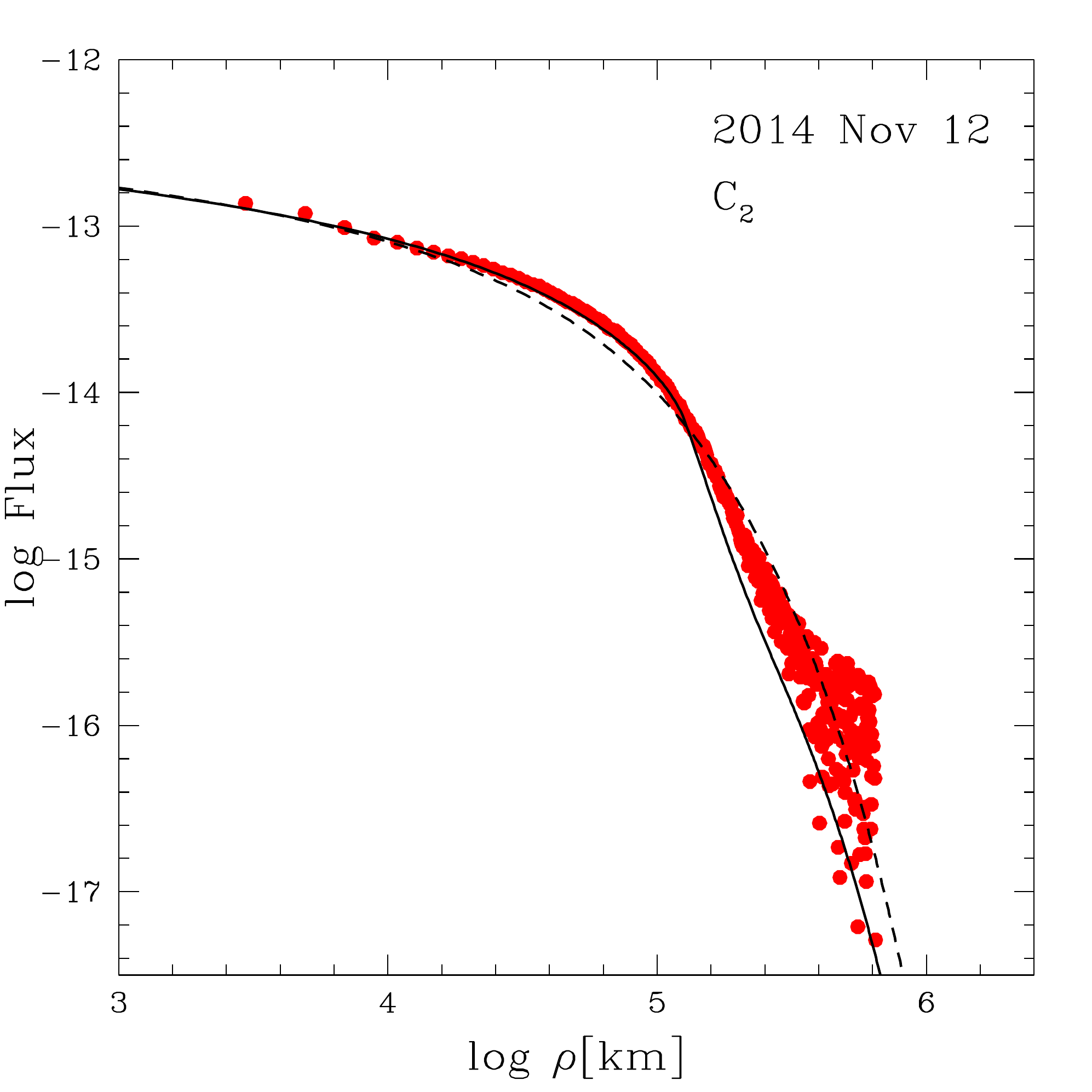}}
\hspace{0.5cm}
\subfigure{ \includegraphics[width=5.5cm]{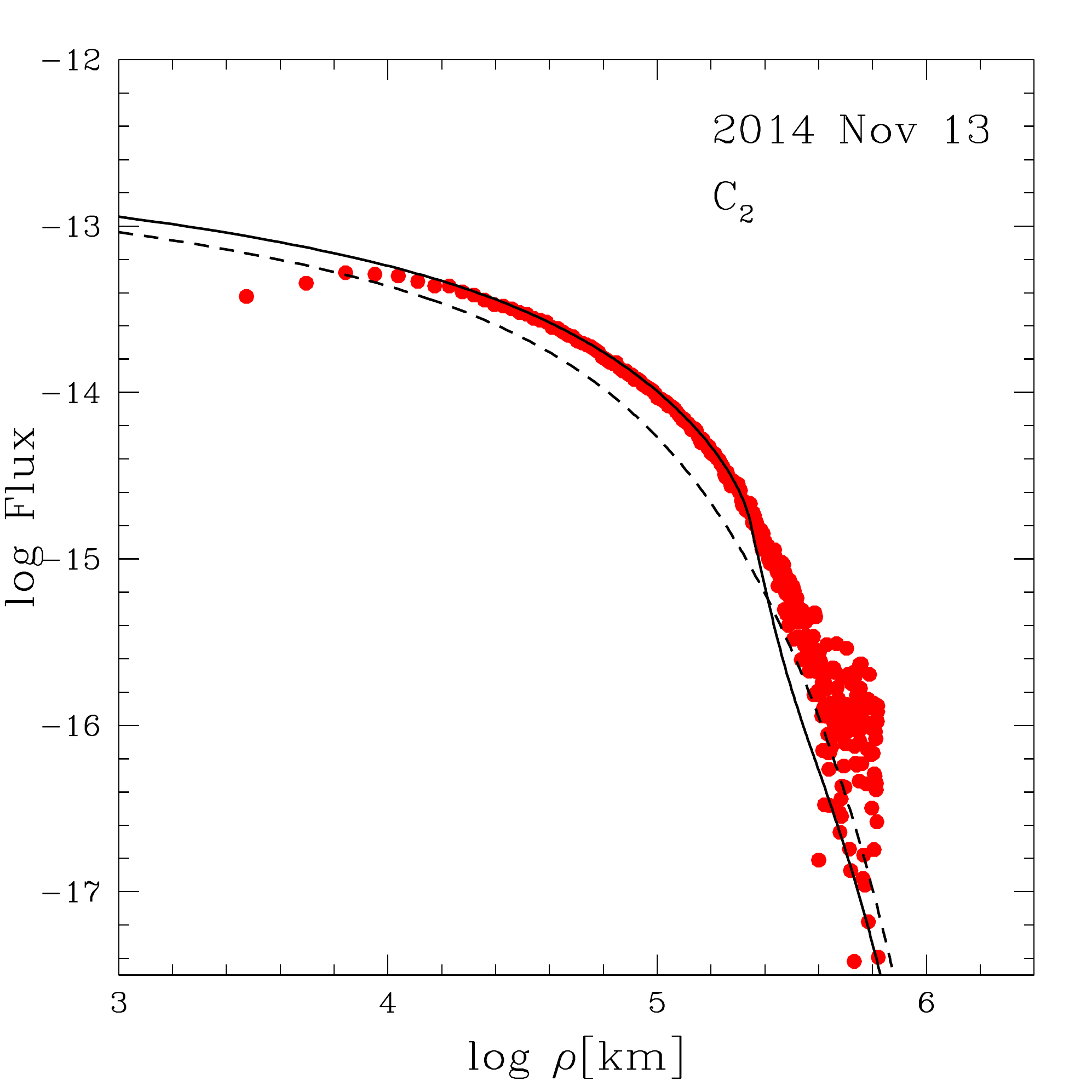}}
\caption{CN (top) and $\mathrm{C_2}$ (bottom) brightness profiles from November 11, 12, and 14 2014 ($r=1.4$ au). The observed profiles are represented with red dots, the Haser model is represented by the dashed line and the outburst model is represented by the black line.}
  \label{profilesoutburst}
\end{figure*}

\indent
In section \ref{rate}, we inferred that Siding Spring underwent an outburst approximately two weeks after its perihelion passage. From our dataset, we can indeed observe an increase of both gas and dust productions  consistent with an outburst occurring between Nov 7.02 and Nov 11.02, 2014. CN and $\mathrm{C_2}$ radial profiles observed on Nov 11 are diverging from stationary profiles. Fig. \ref{profilesoutburst} shows CN and $\mathrm{C_2}$ radial profiles from Nov 11, i.e. our first observation after the outburst, November 12, and November 13. The most prominent feature visible on the Nov 11 CN profile is a bump, with a rather well defined edge at around 60,000~km from the nucleus, overlaid to the usual stationary radial profile which is still visible at large nucleocentric distance (after 125,000~km). In the following two profiles, this feature is moving away from the nucleus as it propagates into the coma. The same feature is also visible in the $\mathrm{C_2}$ profiles. We believe that this is the signature of the outburst. A large amount of gas is released at the time of the outburst, and then expands through the coma. Measuring the position of this shell of material from day to day could in theory allow us to derive the expansion speed of the gas in the coma. However, as can be seen in Fig. \ref{profilesoutburst}, its edge is not sharp enough to determine unambiguously the position of the gas shell in each profile. On Nov 14 and 15, the shape of the CN profile gradually comes back to its pre-outburst shape.

\indent 
We attempted to reproduce the shape of the radial profiles observed in Fig. \ref{profilesoutburst} in order to constrain the ejection velocity of the gas during the outburst, and the time at which it occurred. For each profile in Fig. \ref{profilesoutburst}, the dashed line represents the Haser model adjusted on the data. As expected, the Haser model fails to reproduce the shape of these radial profiles, observed shortly after the outburst. This also means that the production rates we derived in Sect. \ref{rate} are not representative of the real gas production at the time of the outburst. From the profile shape, we estimate that a two components model could reproduce the observations. We thus considered a constant underlying gas emission $Q_b$ in stationary state and described by the Haser model, on which we added the brief emission of a large amount of gas that expands at a given velocity $v_{o}$ into the coma. We used two exponentials to describe the gas release during the outburst: an increasing exponential with a characteristic timescale $\tau_1$ of typically a few hours to describe the onset of outburst activity, and a decreasing exponential with a characteristic timescale $\tau_2$ of typically a few days to describe the slow decrease of the gas emission following the outburst. We consider the start of the outburst $t_o$ as the time at which the peak gas emission $Q_o$ occurs. As for the Haser model, we assumed spherical symmetry for the gas emission, a constant radial velocity $v_o$, and a single step photodissociation of parent molecules into daughter molecules. We used the same scalelengths as for the rest of our analysis. In our model, the radial density distribution of daughter molecules $n(r)$ is thus described by: 
\begin{equation}
n(r) = \frac{Q}{4\pi r^{2}}\left[\frac{\beta_{0}}{\beta_{1}-\beta_{0}} \right] 
\left[epx(-\beta_{0}r) - epx(-\beta_{1}r) \right]  
\end{equation}
with 

\begin{equation}
Q = \frac{Q_{b}}{v} + \frac{Q_{o}}{v_{o}} exp(-\frac{t_{o}-t}{\tau_{1}}) \hspace{1.5cm} \mathrm{if} \hspace{0.5cm} t<t_{o}
\end{equation}

\begin{equation}
Q = \frac{Q_{b}}{v} + \frac{Q_{o}}{v_{o}} exp(-\frac{t-t_{o}}{\tau_{2}}) \hspace{1.5cm} \mathrm{if} \hspace{0.5cm} t>t_{o}
\end{equation}
In these equations $r$ is the radial distance in the coma, $v$ is the velocity of the gas in stationary equilibrium, fixed to 1 km/s as in Sect. \ref{rate}, and $\beta_{0}$ and $\beta_{1}$ are respectively the parent and daughter scalelengths. Firstly, we tried to model the outburst using only one exponential to simulate the decrease of the gas emission directly following the outburst. However, the transition between stationary equilibrium and the outburst was too sharp compared to the profiles shown in Fig. \ref{profilesoutburst} and we added a second exponential to simulate the onset of the activity. We also tried to use simpler functions as slopes but it could not reproduce the shape of the profiles observed at the time of the outburst as well as with exponentials.

\indent 
For CN, we determined the constant gas emission $Q_b$ by adjusting the Haser model on the Nov 11 observations. We made the adjustment between cometocentric distances of 100,000~km and 160,000~km, since the gas released during the outburst could not have had the time to reach this part of the coma yet. We obtained $Q_b(\mathrm{CN}) = 4.2$ $10^{25}$ mol/s. This production rate is close to the one measured four days before, on Nov 7 (see Table \ref{obstabtrappist}). As explained before, we used a constant value of $Q_b$ for all outburst profiles. Even if the gas production rates varies from day to day, as can be seen from Table \ref{obstabtrappist}, this variation remains small compared to the amount of gas released at the time of the outburst. Furthermore, on Nov 12 and 13 profiles, the influence of the gas released during the outburst extends so far in the coma that it is impossible to independently adjust the value of $Q_b$ as was done for the Nov 11 profile. We then tried to constrain the values of $Q_o$, $v_o$, $t_o$, $\tau_1$, and $\tau_2$ that match simultaneously the observed profiles on Nov 11, 12, and 13. From the CN profiles, we obtained $Q_o(CN)= 2.7\pm0.5$ $10^{26}$ mol/s, $v_o= 1.1\pm0.2$ km/s, $\tau_1 = 0.3 \pm 0.1$ j, $\tau_2 = 1.6 \pm 0.2$ j and $t_o=$ Nov 10.6$\pm$0.2. This set of models is overlaid to the observed CN profiles in the upper part of Fig. \ref{profilesoutburst} using a black line. For the three dates, the observed CN profile is much better represented by our model (black curve) than by the Haser model (black dashed line). This model is able to reproduce the bump observed in the radial profile, and its propagation through the coma during the days following the outburst. Larger discrepancies between the observed and model profiles only appear at large nucleocentric distances though. This is mainly caused by the uncertainties in the measurements due to the high airmass of these observations and the faint signal at these distances.

\indent
We repeated the same process for the $\mathrm{C_2}$ profiles. We measured $Q_b(\mathrm{C_2}) = 6.55$ $10^{25}$ mol/s. Once again, this value is close to the one measured on Nov 7. We then attempted to find the best set of parameters to simultaneously reproduce the three $\mathrm{C_2}$ profiles. We obtained $Q_o(\mathrm{C_2})= 3.9\pm0.5$ $10^{26}$ mol/s, $v_o= 1.1\pm0.2$ km/s, $\tau_1 = 0.4 \pm 0.1$ j, $\tau_2 = 1.8 \pm 0.2$ j and $t_o=$ Nov 10.7$\pm$0.2. This set of models is overlaid to the observed $\mathrm{C_2}$ profiles in the lower part of Fig. \ref{profilesoutburst}. For the Nov 11 profile, our model is a much better representation of the data than the Haser model. For the other two profiles, the agreement is less good than for CN, especially at large nucleocentric distances, but still better than with the Haser model. We were not able to apply our model on the OH data because the SNR of these profiles was too poor. We do not have any NH observations at the time of the outburst, and we do not have $\mathrm{C_3}$ observations the day after the outburst either to perform the same analysis. We note that it is usual for a cometary outburst that the typical time to reach the peak emission is shorter than the recovery time, even though the characteristic times for an outburst vary from one comet to another (see \citealt{Schleicher2002}, \citealt{Li2011}, \citealt{Manfroid2007}). The $\mathrm{C_2}$/CN ratio produced during the outburst is 1.46. Within the error bars, this is consistent with the ratios measured before the outburst, indicating that no obvious change of the coma gas composition occurred during the outburst.

\indent
Our model provides a satisfactory representation of the shape of the radial brightness profile of the coma of comet Siding Spring during three days following an outburst, using one single set of parameters for each species. Parameters derived from the $\mathrm{C_2}$ profiles and from the CN profiles are still with each other within the error bars. We can determine that the outburst most probably occurred on Nov 10 around 15:30~UT ($\pm$ 5h). The speed of the gas in the outburst was $v_o= 1.1\pm0.2$ km/s. This velocity is consistent with the gas outflow velocity of 1 km/s that we assumed in the Haser model. Only few measurements of gas velocity during an outbursts have been made, but our value is close to the CN radial expansion velocity of $0.85\pm0.04$ km/s measured by \cite{Schulz2000} during the activity outburst of comet C/1995 O1 (Hale-Bopp). \cite{Manfroid2007} measured a gas velocities of $0.4-0.6$ km/s in the coma of comet 9P/Tempel 1 after the Deep Impact event. This is lower than the velocity we measured here, but the outburst of comet Tempel 1 was caused by the impact of a 362 kg mass on the nucleus and the outburst mechanism is thus different from the one observed here. Measurements of gas outflow velocities in the coma of quiescent comets have been made at various heliocentric distances, and on comets more or less active. The velocity we measure here is in agreement with gas velocities in the coma of comets at the same heliocentric distance (\citealt{Bockelee1990}, \citealt{Biver1997}, \citealt{Tseng2007}, for example), and with HCN shells expansion velocities measured in the coma of comet 8P/Tuttle \citep{Waniak2009}. 

\indent
Unfortunately, we were only able to determine an upper limit for the velocity of the dust ejected during the outburst. We attempted to measure the position of the dust shell ejected during the outburst in successive BC images taken the days after the outburst. However, the SNR in the images is low, making the determination of the dust shell position difficult. The SNR was higher in RC filter but we do not have regular observations in this filter around the time of the outburst. We could only determine an upper limit of 100 m/s. This may seems low compared to the dust velocities usually measured during cometary outbursts. Indeed, \cite{Hsieh2010} and \cite{Lin2009} measured velocities around 550 m/s for the outburst of comet 17P/Holmes, while \cite{Schulz2000} measured a velocity of 200 m/s for the outburst of comet C/1995 O1 (Hale-Bopp), \cite{Lara2007} measured velocities ranging from 150 to 230 m/s in the coma of comet 9P/Tempel 1 after the Deep Impact event. Such a low dust velocity could indicate that the outburst mostly released large grains.


\subsection{Modelling the outburst and activity pattern}
\label{model}

Based on the gas production rates determined in the previous sections, we aim at estimating the thickness of dust that would be required at the surface to reproduce the pre-perihelion production rates, using a similar technique as the one described in \cite{Guilbert2014}, as well as the outburst observed in our dataset. We based our calculation on the measurement of the CN production rates. CN is the daughter molecule of HCN, which if present in the nucleus under a dusty crust, may not sublimate freely. If buried under a dusty crust, it would follow a diffusion regime, with a mass loss rate written as (assuming an ideal gas law, \citet{Fanale1984, Schorghofer2008, Gundlach2011, Guilbert2014}):
\begin{equation}
\label{mpr}
J \sim \frac{\varphi}{\Delta r}~\sqrt{\frac{2m}{\pi k_B T}}~P_S(T),
\end{equation}
with $J$ the ice loss rate from a subsurface layer buried under the crust, $\varphi$ the permeability of the crust, $\Delta r$ [m] its thickness, $m$ [kg] the molecular weight, $k_B$ [JK$^{-1}$] the Boltzmann constant, $T$ [K] the temperature, and $P_S(T)$ [Pa] the saturation vapour pressure, which is given by the Clausius-Clapeyron equation
\begin{equation}
P_S(T) = \alpha ~ e^{-E_a/k_B T}
\end{equation}
with $\alpha=3.86 \times 10^{10}$~Pa and $-E_a/k_B=4023.66$~K for HCN. 
In our calculations, the temperature distribution at the surface and in the nucleus of the comet is determined using a numerical model of three-dimensional heat transport \citep{Guilbert2011}. The model solves the heat equation taking into account conduction via contacts between grains, radiation within pores, insolation, and thermal emission at the surface. This distribution is computed as a function of time and orbital position of the comet. Since Siding Spring is a dynamically new comet, with no previously known perihelion passage, we start our simulation at 100~au with a temperature of 20~K.  We assume that the upper layer is made of porous dust, with 85\% porosity and thermal inertia of 15~$Jm^{-2}K^{-1}s^{-1/2}$, which is consistent with the measurements of low thermal inertias for comets in general and 67P/Churyumov-Gerasimenko, which is being thoroughly studied by the ESA/Rosetta mission \citep{Lowry2012, Leyrat2015}. We then estimate the thickness $\Delta r$ of the dusty crust by inverting Eq.\ref{mpr} for each orbital position of the comet where we have a measurement of CN production rate.

Given that our study of the coma morphology has shown that Siding Spring presents jets, it is unlikely that the entire surface of its nucleus may be active: we expect that this would lead to local variations in the crust thickness. We must therefore keep in mind that our modelling is aimed at providing order of magnitude results, so to understand the general behaviour of Siding Spring's activity. Before the outburst, our calculations show that CN production rates can be reproduced if we assume that the thickness of the dusty crust at the surface of Siding Spring (assuming that the overall surface is active) is steadily increasing, from 3.5~cm on June 03, when at 2.42~au, to $\sim$8~cm on October 29, when at 1.4~au. This is consistent with the expected progressive formation of a dust mantle at the surface of an active comet, and also consistent with our inference of a coma becoming less dusty when the comet approaches the Sun. Eventually, the porosity of the dust mantle may become too small to allow any particle to be released, thus forming a stable crust, with a cohesive strength greater than the vapour pressure building up below it. However, with insolation becoming more efficient in sublimating subsurface ice when the comet reaches perihelion, the gas pressure below the crust would increase, so that the dust mantle may be totally or partially blown off. This process may be invoked to explain the outburst of comet Siding Spring. Indeed, from Nov 7, our calculations show that the crust thickness required to reproduce the CN production rates decreases, with a minimal thickness of 4.5~cm achieved on Nov 11. After that, the crust thickness increases again. It is unclear from the available data whether the whole dust mantle was blown off, or only part of it. Although many parameters involved in these calculations are unknown or poorly constrained, we can estimate that 50 to 120 $\times$ 10$^6$~kg of dust were released during the outburst. These numbers are in fact consistent with the dust production we measured, and the low velocity of dust particles at the time of the outburst which indicates that large grains were emitted. Indeed, the dust particles forming a dust mantle are expected to be the heaviest relative to their cross section to be entrained by the outgassing. Therefore, by basing our calculations on the CN production rates measured before and during the outburst, we can reproduce the activity pattern and dust properties observed for comet Siding Spring in a self-consistent way. This may indicate that the mechanism at the origin of Siding Spring's outburst is related to dust mantle formation and destruction.

\section{Conclusion}
\label{summary}

In this paper, we have presented a unique dataset consisting in more than a year of regular observations of comet C/2013 A1 (Siding Spring) with the TRAPPIST telescope in Chile, along with low-resolution spectra obtained with the ESO/VLT FORS~2 instrument. We first demonstrated that simultaneous measurements of gas radial profiles and production rates from narrow-band photometry with TRAPPIST and low-resolution spectroscopy with FORS~2 were in good agreement and could be used to complement each other in the future. We then studied the evolution of the comet activity, along with its gas and dust composition, using various ratios such as CN/OH, $\mathrm{C_2}$/OH, or $\mathrm{C_3}$/OH, which showed little or no variation with heliocentric distance. Only the NH/OH ratio increased while the comet approached the Sun, but this could just be the effect of poorly known NH scalelengths.
Enhancement techniques applied to all images allowed us to detect morphological features in the coma. Except for the dust and CN, these features were weak and could not be detected in all images. The orientation of $\mathrm{C_3}$ and $\mathrm{C_2}$ features is difficult to determine due to their small spatial extension. CN and dust features both have an hourglass shape, but are not oriented in the same direction. This indicates that they probably do not originate from the same source region(s).
This monitoring allowed us to detect an outburst of activity, which we have characterized and modelled. The CN and $\mathrm{C_2}$ production rates were measured at the peak of the outburst. The $\mathrm{C_2}$/CN ratio measured at that time is consistent within the error bars with the ratios measured before the outburst, while the production rates were multiplied by about a factor 5. Overall, we could not detect any significant change of the coma composition or dust-to-gas ratio after perihelion, nor at the time of the outburst, which may indicate a certain level of homogeneity of the nucleus composition.

The gas radial profiles observed shortly after the outburst were different from those usually observed for this comet. We provided a simple model for representing them. This model represents the outburst as a sudden emission of a large quantity of gas that expands in the coma at a given velocity. The outburst is overlaid to the equilibrium gas emission described by the Haser model. We considered an exponential rise of the gas emission with a characteristic timescale $\tau_1$ to reach a maximum production rate, then an exponential decrease with a characteristic timescale $\tau_2$. For consistency, we used the same hypothesis as in the Haser model, as well as the same scalelengths. We found a set of modelling parameters that simultaneously provided a good representation of the radial profiles for the three days following the outburst. In addition, the parameters determined from CN and $\mathrm{C_{2}}$ profiles agree within the error bars. 
We used a thermal evolution model to reproduce the activity pattern and outburst, by constraining the thickness of a dusty crust present at the surface of the comet. Our results are consistent with the progressive formation of a dust mantle, which may be partially blown off during the outburst. Overall, our observations and modelling results, i.e. the shallow dependence of pre-perihelion gas production rates, consistent with the progressive formation of a dust mantle, are in fact quite typical for dynamically new comets as concluded by \cite{AHearn1995}. Comet Siding Spring is therefore behaving like a typical dynamically new comet.

For example, the evolution of Siding Spring's activity between 2.5~au and perihelion is relatively similar to that of comet C/2009 P1 (Garradd) (\citealt{Combi2013} and \citealt{Bodewits2014}) at around the same pre-perihelion heliocentric distance. The behavior comet Garradd's activity was attributed to the presence of an extended source of water. According to \cite{Bodewits2015}, $\mathrm{CO_{2}}$ sublimation from a constant area on the nucleus may have dominated the activity of comet Siding Spring at heliocentric distances larger than 2.5~au. At this distance, the sublimation of icy grains, in addition to ice sublimating from the nucleus itself, may explain the increase of the water production, as it was the case for comet Garradd. Icy grains in the coma of this comet have also been invoked by \cite{Li2014} to explain the dust color trend in the coma and its temporal evolution.The scenario suggested by \cite{Bodewits2015} thus seems consistent with our results.

Even though outbursts may seem as a rather common phenomena among comets, their origin is still not well understood. The amplitude of the brightness increases and their frequency vary from comet to comet. Some comet nuclei may be splitting: such an event would expose a large amount a fresh ice that would explain the sudden increase of gas and dust production. We carefully searched in the enhanced images of the comet for five days after the outburst and could not detect any sign of nucleus fragmentation, such as a blob of material in the coma, or the so-called "coma wings" \citep{Boehnhardt2000}. A nucleus splitting being the origin of the outburst thus seems unlikely. Outbursts can also be caused by a collision with another body. However, these collisions are extremely rare and no evidence in our observations points to an external source as the origin of this outburst. Cavities under the surface of the nucleus containing gas under high pressure have already been invoked to explain cometary outbursts \citep{Ipatov2011}. This process is similar to the one we have inferred in our model, where a dusty crust may be totally or partially blown off due to an increase of the gas pressure underneath. Although this scenario may adequately reproduce both gas and dust production evolution before and during the outburst, we cannot exclude other mechanisms. In particular, by studying the coma morphology, \cite{Li2014} determined two pole solutions for comet Siding Spring resulting in different seasonal effect on the nucleus. For their second solution (rotational pole: Right Ascension = $190\pm10\degree$ and Declination = $50\pm5\degree$), they predicted that the Sun would illuminate a new area just after perihelion. The onset of sublimation on this fresh newly active area of the nucleus just after perihelion may thus also explain the outburst we observed on November 10.

\begin{acknowledgements}
TRAPPIST is a project funded by the Belgian Fund for Scientific Research (Fond National de la Recherche Scientifique, F.R.S.-FNRS) under grant FRFC 2.5.594.09.F, with the participation of the Swiss National Science Fundation (SNF). C. Opitom acknowledges the support of the FNRS. E. Jehin and M. Gillon are FNRS Research Associates, D. Hutsem\'ekers is FNRS Senior Research Associate and Jean Manfroid is Research Director of the FNRS. We are grateful to David Schleicher and the Lowell Observatory for the loan of a set of NASA HB comet filters.
\end{acknowledgements}


\bibliographystyle{aa}
\bibliography{Biblio}

\end{document}